# Formation of Self-Organized Anode Patterns in Arc Discharge Simulations

**Juan Pablo Trelles**

Department of Mechanical Engineering, University of Massachusetts Lowell, Lowell, MA 01854, USA

E-mail: Juan_Trelles@uml.edu

**Abstract**

Pattern formation and self-organization are phenomena commonly observed experimentally in diverse types of plasma systems, including atmospheric-pressure electric arc discharges. However, numerical simulations reproducing anode pattern formation in arc discharges have proven exceedingly elusive. Time-dependent three-dimensional thermodynamic nonequilibrium simulations reveal the spontaneous formation of self-organized patterns of anode attachment spots in the free-burning arc, a canonical thermal plasma flow established by a constant DC current between an axi-symmetric electrodes configuration in the absence of external forcing. The number of spots, their size, and distribution within the pattern depend on the applied total current and on the resolution of the spatial discretization, whereas the main properties of the plasma flow, such as maximum temperatures, velocity, and voltage drop, depend only on the former. The sensibility of the solution to the spatial discretization stresses the computational requirements for comprehensive arc discharge simulations. The obtained anode patterns qualitatively agree with experimental observations and confirm that the spots originate at the fringes of the arc – anode attachment. The results imply that heavy-species – electron energy equilibration, in addition to thermal instability, has a dominant role in the formation of anode spots in arc discharges.

## 1. Introduction

Pattern formation and self-organization are complex and fascinating phenomena common in diverse types of biological, chemical, and physical systems. Their occurrence has been associated to instability, symmetry breaking, bifurcation, and to the formation of dissipative or coherent structures to maximize entropy production in systems distant from thermodynamic equilibrium [1, 2]. Pattern formation and self-organization are also prevalent in electrical discharges, particularly in the regions near the electrodes, as evidenced by current transfer spots [3, 76, 78].

Electrode spots may produce detrimental effects within plasma processing applications (e.g., break down of process uniformity) or may limit the life of the electrodes (e.g., due to localized heating and subsequent material evaporation). Therefore, the understanding of pattern formation in electrical discharges is of interest not only from a fundamental, but also from practical and technological, point of view.

Different types of pattern formation phenomena have been reported in a wide range of electrical discharges, from low-pressure – high-current vacuum arcs [4, 5], to low-pressure – low-current glow [6], streamer [7], and dielectric barrier (DBD) [8] discharges, to high-pressure – low-current glow [9, 10, 32], DBD [11, 12], and arc [13] discharges, and to high-pressure – high-current arc discharges [14, 15, 16, 17]. The phenomenological behaviors associated to the formation of electrode patterns are significantly different among these discharges. For example, for high current vacuum arcs, as the total current is increased, the anode attachment transitions from diffuse to constricted, possibly due to the combination of magnetic constriction and electrode material evaporation effects [4]. In contrast, for high-pressure low-current glow discharges, as the current is increased, the structures of the anode spots transition from a constricted homogeneous spot to a pattern consisting of small distinct spots, probably due to the intrinsic electrical characteristics of the system [10].

The understanding of pattern formation in atmospheric-pressure arcs is more limited than for other types of plasmas, probably due to the harsh conditions typical in these discharges (e.g., temperatures above $10^4$ [K], current densities greater than $10^5$ [A-m$^{-2}$], and heat fluxes of more than $10^4$ [kW-m$^{-2}$] [28]), which limit the type and extent of diagnostics that can be performed, combined to the complex interaction of fluid dynamic, thermal, chemical, and electromagnetic phenomena [18]. This limited understanding is evidenced by the scarce scientific literature about the topic or by inconsistent or incomplete research findings. For example, the authors of the experimental investigation reported in [14] indicate that the transition among multiple anode attachment spots in the free-burning arc occurs due to the development of a thermal instability between Joule heating and gas cooling, whereas the authors of the comprehensive instability analysis in [17] and experiments reported in [16], conclude that multiple anode attachments form only when both, the electron overheating instability and the evaporation–ionization instability, are active. Furthermore, the experimental and theoretical studies in [19] show that the arc attachments on cold (passive) and hot (active) anodes are very different due to the dominance of different modes of instability; therefore, it could be expected that the instability dominating the attachment of an arc to a strongly cooled anode cannot be unambiguously identified.

Computational modeling and simulation provides a new dimension, in addition to experimental and theoretical approaches, for the investigation of electrode pattern formation in electrical discharges. The success of computational research has been markedly different between high-pressure arc and other types of discharges. In low-pressure – low-current discharges, drift-diffusion models have proven successful to describe the incidence of pattern formation in DBD and streamer discharges [7], and even the self-organization of one-, two-, and three-dimensional patterns in glow discharges [33, 34]. In addition, time-dependent two-dimensional (axi-symmetric) drift-diffusion nonequilibrium models have been able to capture the spontaneous occurrence of anode spots in high-pressure glow discharges [20]. These computational studies led to the understanding that the origin for pattern formation in those discharges is rooted in charge transport – electrodynamic, rather than thermal, phenomena.

In contrast, computational investigations of spontaneous electrode pattern formation in high-pressure – high-current arc discharges have been scarcer. This may be caused by, from one end, the strong dependence on fluid dynamic effects, in addition to the electrodynamic and reactive effects that dominate low-pressure low-current discharges, which drastically increases the complexity of computational analyses; and from the other, due to the lack of widespread use of thermodynamic nonequilibrium arc discharge models, such as those reported in [21, 22, 23], which provide a more adequate description of the interactions between the bulk plasma and the electrodes. Additionally, the physical processes responsible for the formation of patterns are markedly different between cathodes and anodes. Whereas cathode patterns can be investigated by partially isolating the cathode region, to adequately describe the anode region, not only the electrode, but also the adjacent plasma, needs to be analyzed. For example, the modeling of thermionic cathodes in [24] reveals the set of possible current transfer spot patterns, and points to the fact that the question of stability of steady-state solutions is still open. Comparable analyses for the anode region, to be best knowledge of the authors, have not been reported to date. The report by Benilov [25] provides a comprehensive summary of the modeling of electrode regions in high-pressure arc discharges, and the review by Heberlein *et al* [75] delivers a detailed summary of the current understanding of the anode region of electric arcs.

Preliminary numerical investigations relevant to anode spot formation in arc discharges may be traced back to the studies by Amakawa *et al* [26] of the anode region of an atmospheric-pressure arc subject to an impinging flow vertical to the anode surface using a steady-state two-dimensional thermodynamic nonequilibrium model. The results captured the experimentally observed diffuse or constricted anode attachments for high and low flow rates, respectively. Nevertheless, convergent numerical solutions were only achievable for low or high flow rates, but not for intermediate ones, which may be indicative to an underlying transition phenomenon that could not be captured by a steady-state flow model. It could arguably be expected that the use of a time-dependent model, together with robust and high accuracy numerical methods, may lead to the capturing of spontaneous anode attachment spot formation, in a similar manner as in the DBD simulations in [20], for the intermediate flow rates. Additional insight into the role of time-dependent and three-dimensional numerical models to describe an apparently stead-state axi-symmetric flow is provided by the study conducted by Kaddani *et al* [27] of a free-burning arc using a LTE model. Their results indicate that instabilities inherently develop in a transient and three-dimensional model, which would otherwise be mitigated by the forced symmetry in two-dimensional or steady-state models (e.g., the results reported in [74] using a 3D steady-state model did not report the occurrence of anode patterns).

This article reports for the first time the computational investigation of the spontaneous formation of anode attachment spot patterns in high-pressure high-current arc discharge, i.e., the atmospheric-pressure free-burning arc. The free-burning arc is a canonical thermal plasma flow in which an electric arc, formed by a constant DC current between a conical cathode aligned vertically on top of a flat horizontal anode in the absence of external forcing (e.g., auxiliary gas flow, imposed magnetic field), establishes a recirculating flow of gas and a corresponding bell-shaped optical emission pattern from the plasma [28]. Due to its relative simplicity and relevance in diverse technological applications (e.g., arc welding, metallurgy, circuit breakers), the free-burning arc has been extensively studied, both experimentally and computationally, and is commonly used for benchmarking computational plasma flow models [29, 27, 60-71, 73, 74].

Given the geometrical symmetry and constancy of operating conditions, the free-burning arc is usually modeled using steady-state and axi-symmetric flow models. Furthermore, thermal plasmas are traditionally described using models based on the Local Thermodynamic Equilibrium (LTE) assumption. The energy of heavy-species (molecules, atoms, ions) and the free electrons in a plasma in LTE, due to the high collision frequencies among its constituent particles, can be characterized by a single temperature [28]. The LTE approximation is largely valid in the core of the plasma, but it is often invalid when the plasma interacts with another medium, such as solid electrodes or a surrounding cold gas. In this computational research, the free-burning arc is described using a time-dependent three-dimensional thermodynamic nonequilibrium (i.e., non-LTE or NLTE or two-temperature) model. The NLTE model describes the evolution and interaction of the heavy-species and electron temperatures using different energy conservation equations for the heavy-species and electrons, respectively. The model relies on the chemical equilibrium assumption, uses relatively simple boundary conditions over the anode surface (e.g., non-slip flow, convective heat transfer and constant electric potential), and does not include electrode sheath models, charge separation, modeling of the bulk electrodes, ambipolar diffusion effects, detailed radiative heat transfer, or anode material evaporation effects [62-69, 76].

The modeling results reveal the spontaneous formation of anode attachment spots patterns in qualitative agreement with experimental observations in water-cooled metal anodes [17, 14]. These results corroborate the statement in [25] indicating that the adequate simulation of plasma-anode interactions requires the coupled modeling of not only the near-anode nonequilibrium layer, but also of the adjacent bulk plasma. Nevertheless, the simplified nonequilibrium model used in the present work offer a complementary view of the physics dominating anode attachment formation. The obtained results reported in this article indicate that numerical simulations without artificial artifacts can capture the formation of anode spots in atmospheric-pressure arcs, that the characteristics of the spots (e.g., number, location, size) depend on the numerical characteristics of the simulation (e.g., spatial grid resolution), and point towards the need for the use of thermodynamic nonequilibrium models for the comprehensive description of arc discharges.

The paper is organized as follows: Section 2 presents the mathematical plasma flow model based on a fully-coupled monolithic treatment of the thermodynamic nonequilibrium and chemical equilibrium fluid model together with the electromagnetic field evolution equations. Section 3 describes the numerical method used based on the Variational Multiscale Finite Element Method [30, 31], and implemented in a time-implicit second-order-accurate in time and space discretization approach. Section 4 describes the free-burning arc problem: the geometry of the spatial domain, the computational discretization, and the boundary conditions used. Section 5 presents the computational results for the free-burning arc operating with argon for different values of total current. The summary and conclusions of the research are presented in Section 6.

**2. Mathematical Model**

The formulation of the mathematical arc discharge model is composed of three parts: a fluid flow model, an electromagnetic field evolution model, and models for the material properties and constitutive relations.

*2.1. Fluid flow*

The plasma is described as a compressible, reactive, electromagnetic fluid in chemical equilibrium and thermodynamic nonequilibrium (NLTE). The fluid model is justified by the relatively high collision frequencies, and therefore short mean-free-paths, among the constituent particles in high-pressure arc discharge plasmas. The flow is described by the set of conservation equations for: (1) total mass, (2) mass-averaged linear momentum, (3) thermal energy of heavy-species, and (4) thermal energy of electrons. These equations are summarized in Table 1 as a single set of *transient – advective – diffusive – reactive* (TADR) transport equations.

**Table 1.** Fluid conservation equations for the NLTE plasma flow model; for each conservation equation: *Transient + Advective – Diffusive – Reactive* = 0.

| Conservation | Transient | Advective | Diffusive | Reactive |
|---|---|---|---|---|
| Total mass | $\partial_t \rho$ | $\nabla \cdot (\mathbf{u}\rho)$ | 0 | 0 |
| Mass-averaged momentum | $\partial_t \rho \mathbf{u}$ | $\nabla \cdot (\mathbf{u} \otimes \mathbf{u}\rho + p\boldsymbol{\delta})$ | $-\nabla \cdot \boldsymbol{\tau}$ | $\mathbf{J}_q \times \mathbf{B}$ |
| Internal energy heavy-species | $\partial_t \rho h_h$ | $\nabla \cdot (\mathbf{u}\rho h_h)$ | $-\nabla \cdot \mathbf{q}'_h$ | $D_t p_h + S_{eh} - \boldsymbol{\tau} : \nabla \mathbf{u}$ |
| Internal energy electrons | $\partial_t \rho h_e$ | $\nabla \cdot (\mathbf{u}\rho h_e)$ | $-\nabla \cdot \mathbf{q}'_e$ | $D_t p_e - S_{eh} - S_r + \mathbf{J}_q \cdot (\mathbf{E} + \mathbf{u} \times \mathbf{B})$ |

In Table 1, $\partial_t \equiv \partial / \partial t$ is the partial derivative with respect to time, $\nabla$ and $\nabla \cdot$ are the gradient and the divergence operators, respectively, $\rho$ represents total mass density, $\mathbf{u}$ mass-averaged velocity, $p$ total pressure, $\boldsymbol{\delta}$ the Kronecker delta tensor, $\boldsymbol{\tau}$ the stress tensor, $\mathbf{J}_q$ electric current density, $\mathbf{B}$ magnetic field, and $\mathbf{J}_q \times \mathbf{B}$ the Lorentz force; $h_h$ and $h_e$ are the enthalpy of the heavy-species and electrons, respectively; $\mathbf{q}'_h$ and $\mathbf{q}'_e$ represent total heat flux due to heavy-species and electrons, respectively; $D_t \equiv \partial_t + \nabla \cdot$ is the material derivative, and $p_h$ and $p_e$ the heavy-species and electron pressure, respectively; $S_{eh}$ is the electron – heavy species energy exchange term, and $S_r$ represents the volumetric net radiation losses; the term $-\boldsymbol{\tau} : \nabla \mathbf{u}$ represents viscous dissipation and $\mathbf{J}_q \cdot (\mathbf{E} + \mathbf{u} \times \mathbf{B})$ Joule heating.

The stress tensor $\boldsymbol{\tau}$ describes the diffusive transport of linear momentum and is modeled as for a Newtonian fluid according to [35]:

$$\boldsymbol{\tau} = -\mu(\nabla \mathbf{u} + \nabla \mathbf{u}^\mathrm{T} - \tfrac{2}{3}\mu(\nabla \cdot \mathbf{u})\boldsymbol{\delta}) \quad , \tag{1}$$

where $\mu$ is the dynamic viscosity, the superscript $^\mathsf{T}$ indicates the transpose operator, and the $\frac{2}{3}$ factor next to the fluid dilatation term $\nabla \cdot \mathbf{u}$ arises from the use of Stoke's hypothesis for the dilatational viscosity.

The total heat fluxes $\mathbf{q}'_h$ and $\mathbf{q}'_e$ describe the total amount of energy transported by diffusive processes and are given by:

$$\mathbf{q}'_h = -\kappa_h \nabla T_h + \sum_{s \neq e} h_s \mathbf{J}_s \text{ and} \tag{2}$$

$$\mathbf{q}'_e = -\kappa_e \nabla T_e + h_e \mathbf{J}_e, \tag{3}$$

where $T_h$ and $T_e$ are the heavy-species and electron temperatures, respectively; $\kappa_h$ and $\kappa_e$ the heavy-species and the electron translational thermal conductivities, respectively; $\mathbf{J}_s$ and $h_s$ stand for the diffusive mass transport flux and specific enthalpy of species $s$, respectively; and the summation in Eq. 2 runs over all the heavy-species in the plasma (i.e., all species except electrons). The first term in Eq. 2 and in Eq. 3 represent the heat transported by conduction; whereas the second term the enthalpy transported by mass diffusion.

The description of mass diffusion processes under high-temperature nonequilibrium plasma conditions is very involved due to their dependence on concentration, pressure, and temperature gradients and electromagnetic fields as driving forces [36]. Based on the fact that, due to the chemical equilibrium assumption the plasma composition is a function of its thermodynamic state only (e.g., $p$, $T_h$, $T_e$), the second term in Eq. 2 can be approximated as:

$$\sum_{s \neq e} h_s \mathbf{J}_s \approx -\kappa_r \nabla T_h, \tag{4}$$

where $\kappa_r$ is the so-called *reactive* thermal conductivity, which is a function of $p$, $T_h$, $T_e$; and therefore can be treated as any other material property [37]. Using Eq. 4, Eq. 2 can be stated as:

$$\mathbf{q}'_h = -\kappa_{hr} \nabla T_h, \tag{5}$$

where $\kappa_{hr} = \kappa_h + \kappa_r$ represents the *translational-reactive* heavy-species thermal conductivity.

Considering that the electric current density is dominated by the transport of charge by electrons (due to their smaller mass and higher mobility), the energy transported by electron mass diffusion in Eq. 3 can be approximated by:

$$\mathbf{J}_e \approx -\frac{m_e}{e} \mathbf{J}_q, \tag{6}$$

where $e$ is the elementary electric charge and $m_e$ the electron mass. Using the approximation given by Eq. 6, Eq. 3 is given by:

$$\mathbf{q}'_e = -\kappa_e \nabla T_h - \frac{h_e m_e}{e} \mathbf{J}_q. \tag{7}$$

*2.2. Electromagnetic field*

The plasma is assumed to be a non-relativistic, non-magnetic, and quasi-neutral fluid. Based on these assumptions, the evolution of the electromagnetic field associated to the plasma can be described by the set of macroscopic Maxwell's equations listed in Table 2.

**Table 2.** Electromagnetic equations for the NLTE plasma flow model.

| Name | Equation |
|---|---|
| Ampere's law: | $\nabla \times \mathbf{B} = \mu_0 \mathbf{J}_q$ |
| Faraday's law: | $\nabla \times \mathbf{E}_p = -\partial_t \mathbf{B}$ |
| *Generalized* Ohm's law: | $\mathbf{J}_q = \sigma(\mathbf{E}_p + \mathbf{u} \times \mathbf{B})$ |
| Gauss's law: | $\nabla \cdot \mathbf{J}_q = 0$ |
| Solenoidal Constraint: | $\nabla \cdot \mathbf{B} = 0$ |

In Table 2, $\mu_0$ represents the permeability of free space, $\sigma$ electrical conductivity, and $\mathbf{E}_p$ the *effective* electric field used to describe *generalized* Ohm's laws. Generalized Ohm's laws typically account for Hall effects and provide a more detailed description of charge transport due to diffusion processes. Assuming that electron diffusion provides the major modification to the real electric field $\mathbf{E}$, the effective field $\mathbf{E}_p$ is expressed by:

$$\mathbf{E}_p \approx \mathbf{E} + \frac{\nabla p_e}{e n_e}, \tag{8}$$

where $p_e$ is the electron pressure and $n_e$ the number density of electrons. (Terms accounting for the transport of charge by ion diffusion and Hall effects have been neglected in Eq. 8.) It can be noticed that the Joule heating term in Table 1 involves the *real* electric field ($\mathbf{E}$) and not the *effective* one ($\mathbf{E}_p$). It is customary to assume $\mathbf{E}_p \approx \mathbf{E}$ in LTE models (and therefore neglect the effect of the second term in the right side of Eq. 8).

A particularly useful representation of Maxwell's equations for plasma flow modeling is given by the use of the electromagnetic potentials $\phi_p$ and $\mathbf{A}$, namely the *effective* electric potential and the magnetic vector potential, respectively. These potentials are defined by:

$$\mathbf{E}_p = -\nabla \phi_p - \partial_t \mathbf{A} \text{ and} \tag{9}$$

$$\nabla \times \mathbf{A} = \mathbf{B}. \tag{10}$$

The use of $\phi_p$ and $\mathbf{A}$ allows the *a priori* satisfaction of the solenoidal constraint $\nabla \cdot \mathbf{B} = 0$. Using Eq. 9, Eq. 10, and the Coulomb gauge condition $\nabla \cdot \mathbf{A} = 0$ to define $\mathbf{A}$ uniquely, the set of Maxwell's equations in Table 2 can be expressed by the set of equations of: (1) charge conservation and (2) magnetic induction. The former equation restates Gauss's law, whereas the latter combines Faraday's and Ampere's laws. These equations are listed in Table 3 as a set of TADR equations, similarly as the equations in Table 1.

**Table 3.** Electromagnetic field evolution equations in terms of electromagnetic potentials; for each equation: *Transient + Advective – Diffusive – Reactive* = 0.

| Equation | Transient | Advective | Diffusive | Reactive |
|---|---|---|---|---|
| Conservation charge | 0 | 0 | $\nabla \cdot \sigma(\nabla \phi_p + \partial_t \mathbf{A} - \mathbf{u} \times (\nabla \times \mathbf{A}))$ | 0 |
| Magnetic induction | $\mu_0 \sigma \partial_t \mathbf{A}$ | $\mu_0 \sigma(\nabla \phi_p - \mathbf{u} \times (\nabla \times \mathbf{A}))$ | $\nabla^2 \mathbf{A}$ | **0** |

*2.3. Nonequilibrium plasma flow model*

The set of equations in Table 1 and Table 3 constitute the system of equations describing the evolution of the NLTE plasma flow. The equations in Table 1 are expressed in the so-called (quasi-) *conservative* form. (A transport equation in conservative form can be expressed as $\partial_t \zeta + \nabla \cdot \mathbf{f}_\zeta = 0$ for some conserved quantity $\zeta$ and its total flux $\mathbf{f}_\zeta$.) If total mass conservation (i.e., $\partial \rho + \nabla \cdot (\rho \mathbf{u}) = 0$) is invoked *a priori*, the equations in Table 1 can be expressed in the so-called *advective* form, which is consistent with the form of the electromagnetic equations in Table 3. The final set of fluid – electromagnetic TADR equations describing the nonequilibrium plasma flow model is listed in Table 4.

In Table 4, the viscous heating term $-\boldsymbol{\tau}:\nabla \mathbf{u}$ in the heavy-species energy conservation equation has been omitted because it is typically negligible in the plasma flows of interest; the electron - heavy species energy exchange term is modeled as $S_{eh} = K_{eh}(T_e - T_h)$, where $K_{eh}$ is the electron - heavy species energy exchange coefficient (e.g., $K_{eh}$ is inversely proportional to a characteristic time for energy exchange); the radiation losses are modeled as $S_r = 4\pi \varepsilon_r$, where $\varepsilon_r$ is the effective net emission coefficient; and the charge conservation equation assumes that $\nabla \cdot (\sigma \partial_t \mathbf{A}) \approx 0$, as this term is negligible in the flows of interest and its omission greatly simplifies the implementation of the model by avoiding mixed spatial-temporal derivatives.

The complete system of equations in Table 4 is treated in a fully-coupled monolithic manner as a single TADR transport system. This system is expressed in residual form as:

$$\mathcal{R}(\mathbf{Y}) = \underbrace{\mathbf{A_0}\partial_t \mathbf{Y}}_{\text{transient}} + \underbrace{(\mathbf{A}_i \partial_i)\mathbf{Y}}_{\text{advective}} - \underbrace{\partial_i(\mathbf{K}_{ij}\partial_j \mathbf{Y})}_{\text{diffusive}} - \underbrace{(\mathbf{S_1Y} - \mathbf{S_0})}_{\text{reactive}} = \mathbf{0}, \quad (11)$$

where $\mathcal{R}$ represents the residual of the system of equations, $\mathbf{Y}$ is the vector of unknowns, the sub-indexes $i$ and $j$ stand for each spatial coordinate (e.g., for three-dimensional Cartesian coordinates, $i, j = x, y, z$), and the Einstein summation convention of repeated indexes is used (e.g., $\nabla \cdot \mathbf{a} \equiv \partial_i a_i$). The matrices $\mathbf{A_0}$, $\mathbf{A}_i$, $\mathbf{K}_{ij}$ and $\mathbf{S_1}$ are denoted as the *transient*, *advective*, *diffusive*, and *reactive* (TADR) transport matrices, respectively; which, given the non-linear nature of the model, are functions of $\mathbf{Y}$.

**Table 4.** Set of fluid – electromagnetic equations for the arc discharge flow model; for each equation: *Transient + Advective – Diffusive – Reactive = 0*.

| Equation | Transient | Advective | Diffusive | Reactive |
|---|---|---|---|---|
| Conservation total mass | $\partial_t \rho$ | $\mathbf{u}\cdot\nabla\rho + \rho\nabla\cdot\mathbf{u}$ | 0 | 0 |
| Conservation momentum | $\rho\partial_t\mathbf{u}$ | $\rho\mathbf{u}\cdot\nabla\mathbf{u} + \nabla p$ | $\nabla\cdot\mu(\nabla\mathbf{u}+\nabla\mathbf{u}^T) - \nabla\cdot(\frac{2}{3}\mu(\nabla\cdot\mathbf{u})\boldsymbol{\delta})$ | $\mathbf{J}_q \times \mathbf{B}$ |
| Energy heavy-species | $\rho\partial_t h_h$ | $\rho\mathbf{u}\cdot\nabla h_h$ | $\nabla(\kappa_{hr}\nabla T_h)$ | $\partial_t p_h + \mathbf{u}\cdot\nabla p_h + K_{eh}(T_e - T_h)$ |
| Energy electrons | $\rho\partial_t h_e$ | $\rho\mathbf{u}\cdot\nabla h_e$ | $\nabla(\kappa_e \nabla T_e)$ | $\partial_t p_e + \mathbf{u}\cdot\nabla p_e - K_{eh}(T_e - T_h) - 4\pi\varepsilon_r + \mathbf{J}_q\cdot(\mathbf{E}+\mathbf{u}\times\mathbf{B}) + \frac{5k_B}{2e}\mathbf{J}_q\cdot\nabla T_e$ |
| Conservation charge | 0 | 0 | $\nabla\cdot(\sigma\nabla\phi_p) - \nabla\cdot(\sigma\mathbf{u}\times(\nabla\times\mathbf{A}))$ | 0 |
| Magnetic induction | $\mu_0\sigma\partial_t\mathbf{A}$ | $\mu_0\sigma\nabla\phi_p - \mu_0\sigma\mathbf{u}\times(\nabla\times\mathbf{A})$ | $\nabla^2\mathbf{A}$ | 0 |

Up to this point, any independent set of variables can be chosen as components of the vector **Y**. In the present study, the vector **Y** is chosen as the set of *primitive* variables, i.e.,

$$\mathbf{Y} = [\ p \quad \mathbf{u}^\top \quad T_h \quad T_e \quad \phi_p \quad \mathbf{A}^\top\ ], \tag{12}$$

which is robust for the description of both, incompressible and incompressible flows (e.g., $\rho$ could be used as independent variable instead of $p$, but its behavior is not well defined in the incompressible flow limit [38]), and allows the greatest solution accuracy for the variables of interest (e.g., the solution procedure aims to attain convergence of the heavy-species energy conservation equation directly in terms the variable $T_h$, which is the main variable of interest to describe heavy-species energy, instead of solving for $h_h$ and then finding $T_h$ in an intermediate or post-processing step).

Once the vector **Y** is defined, the transport matrices can be expressed completely in terms of **Y**. For example, the transient term in Eq. 11 for variable $p$, i.e., $\partial_t \rho$, is given by:

$$\partial_t \rho = (\frac{\partial \rho}{\partial p}) \partial_t p + (\frac{\partial \rho}{\partial T_h}) \partial_t T_h + (\frac{\partial \rho}{\partial T_e}) \partial_t T_e, \tag{13}$$

where the first term in the right-hand-side describes acoustic propagation (e.g., negligible or ill defined in incompressible flows), and the second and third terms, the dependence of mass density in heavy-species and electron temperatures, responsible for heat wave expansion.

Given the set of plasma flow equations in Table 4 and the set of independent variable in Eq. 12, closure of the mathematical model requires the definition of thermodynamic (i.e., $\rho$, $\partial \rho / \partial p$, $\partial \rho / \partial T_h$, $\partial \rho / \partial T_e$, $h_h$, $\partial h_h / \partial p$, $\partial h_h / \partial T_h$, etc.) and transport (i.e., $\mu$, $\kappa_{hr}$, $\kappa_e$, $\sigma$) material properties, as well as the terms $K_{eh}$ and $\varepsilon_r$. The explicit form of the TADR matrices used is listed in Appendix A.

*2.4. Material properties and constitutive relations*

The calculation of material properties for a plasma in chemical equilibrium is composed of three consecutive steps: (1) calculation of the plasma composition, (2) calculation of thermodynamics properties, and (3) calculation of transport properties.

The plasma composition can be determined by: the mass action law (minimization of Gibbs free energy), Dalton's law of partial pressures, and the quasi-neutrality condition [28]. The present study considers a four-species argon plasma, composed of the species: $Ar$, $Ar^+$, $Ar^{++}$, and $e^-$ (i.e., argon atoms, single ionized ions, doubly ionized ions, and free electrons). For such plasma, the set of equations to be solved to determine the number density $n_s$ of each species $s$ are:

$$\frac{n_{e^-} n_{Ar^+}}{n_{Ar}} = \frac{Q_{e^-} Q_{Ar^+}}{Q_{Ar}} \left(\frac{2\pi m_e T_e}{h_P^2}\right)^{\frac{3}{2}} \exp\left(-\frac{\varepsilon_{Ar^+}}{k_B T_e}\right), \tag{14}$$

$$\frac{n_{e^-} n_{Ar^{++}}}{n_{Ar^+}} = \frac{Q_{e^-} Q_{Ar^{++}}}{Q_{Ar^+}} \left(\frac{2\pi m_e T_e}{h_P^2}\right)^{\frac{3}{2}} \exp\left(-\frac{\varepsilon_{Ar^{++}}}{k_B T_e}\right), \tag{15}$$

$$n_{Ar} + n_{Ar^+} + n_{Ar^{++}} + \theta n_{e^-} = \frac{p}{k_B}, \text{ and} \tag{16}$$

$$n_{Ar^+} + n_{Ar^{++}} - n_{e^-} = 0; \tag{17}$$

where $k_B$ is Boltzmann's constant, $h_P$ is Planck's constant, $Q_s$ and $\varepsilon_s$ are the partition function and formation (ionization) energy of species $s$, respectively, and obtained from [28, 39]; and $\theta = T_e/T_h$ is the so-called thermodynamic nonequilibrium parameter. Equations 14 and 15 are Saha equations appropriate for the NLTE model (other alternatives have been reported in the literature, e.g. [39]) in which the lowering of the ionization energy has been neglected. Solution of Eqs. 14 to 17 is accomplished using a Newton Method and provides the composition of the plasma in terms of the number densities of each species $n_s$ for given values of $p$, $T_h$, and $T_e$.

Once the plasma composition is known, the thermodynamic properties $\rho$, $h_h$, and $h_e$ are calculated by:

$$\rho = \sum_s m_s n_s, \tag{18}$$

$$h_h = \rho^{-1} \sum_{s \neq e} \left(\tfrac{5}{2} k_B n_s T_h + n_s \varepsilon_s + k_B n_s T_e \frac{dQ_s}{d\ln T_e}\right), \text{ and} \tag{19}$$

$$h_e = \rho^{-1} \tfrac{5}{2} k_B n_{e^-} T_e, \tag{20}$$

where $m_s$ is the mass of species $s$. These properties are depicted in Fig. 1 as function of $T_e$ for different values of $\theta = T_e/T_h$ and for $p = 1$ [atm], where the marked nonlinearity of these properties can be observed. The partial pressures $p_e$ and $p_h$ in the reactive terms of the energy conservation equations in Table 4 are calculated by: $p_e = k_B n_{e^-} T_e$ and $p_h = k_B (n_{Ar} + n_{Ar^+} + n_{Ar^{++}}) T_h$, respectively.

The derivatives of thermodynamic properties required for the TADR model (e.g., Eq. 13 and TADR matrices in Appendix A) are calculated using a finite difference approximation, e.g.,

$$\frac{\partial \rho}{\partial p}(p, T_h, T_e) \approx \frac{\rho(p+\delta p, T_h, T_e) - \rho(p, T_h, T_e)}{\delta p}, \tag{21}$$

where $\rho(p, T_h, T_e)$ explicitly indicates the functional dependence of $\rho$ on $p$, $T_h$, $T_e$, and $\delta p$ is a small discrete differential set equal to 10 [Pa]. Derivatives with respect to $T_h$ and $T_e$ are calculated similarly using $\delta T_h = \delta T_e = 10$ [K]. These values of discrete differentials have been chosen to calculate the derivatives with high accuracy and smoothness (i.e., without numerical noise).

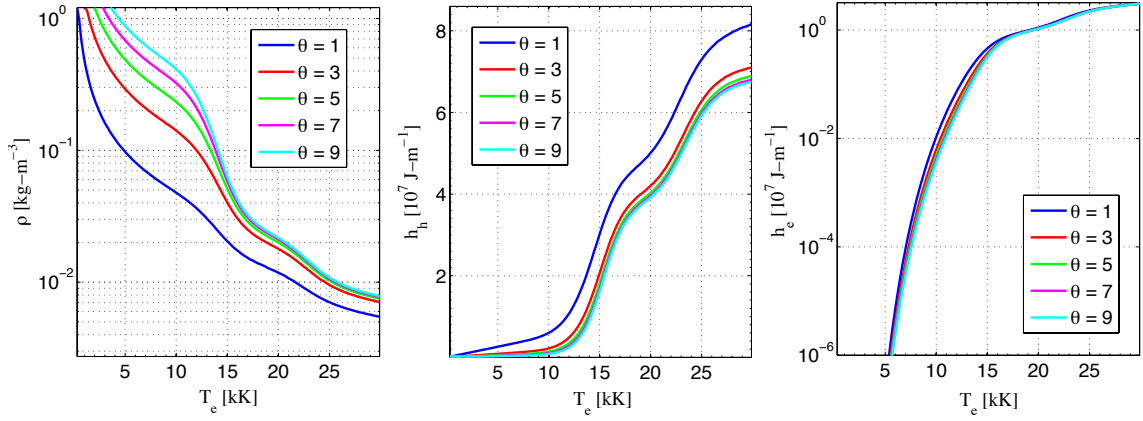

**Figure 1.** Thermodynamic properties: mass density ($\rho$), heavy-species and electron enthalpy ($h_h$ and $h_e$, respectively) for an argon plasma at $p = 1$ [atm] as function of electron temperature $T_e$ and for different values of the nonequilibrium parameter $\theta = T_e/T_h$.

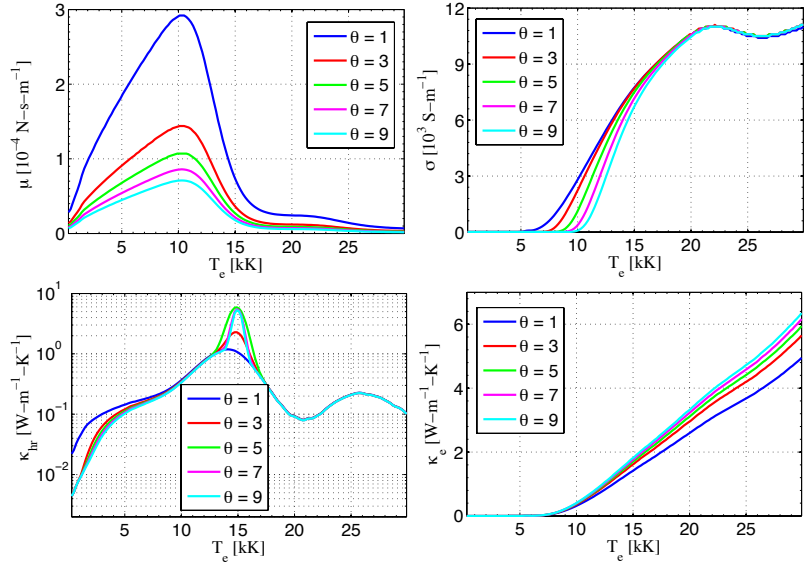

**Figure 2.** Transport properties: viscosity ($\mu$), electrical conductivity ($\sigma$), heavy-species translational – reactive thermal conductivity ($\kappa_{hr}$), and electron thermal conductivity ($\kappa_e$), for an argon plasma at $p = 1$ [atm] as function of $T_e$ for different values of $\theta$.

The accurate calculation of nonequilibrium transport properties for a plasma following the Chapman-Enskog procedure [40, 35] is computationally demanding, particularly within time-dependent three-dimensional simulations. To reduce the computational cost, the NLTE model for the 4-component argon plasma uses look-up tables based on the nonequilibrium transport properties at $p = 1$ [atm] reported in [41, 29]. Figure 2 depicts the transport properties used. Similarly to other electron properties, $\sigma$ and $\kappa_e$ are essentially zero at low temperatures, which drastically increases the stiffness of the numerical solution of Eq. 11.

The net radiation losses $S_r = 4\pi\varepsilon_r$ were modeled using the values of the net emission coefficient $\varepsilon_r$ as function of $T_e$ for an optically thin argon plasma reported in [43] within a table look-up procedure. The volumetric electron – heavy species energy exchange term $S_{eh} = K_{eh}(T_e - T_h)$ models the kinetic equilibration processes between electrons and heavy-species, i.e., the average exchange of kinetic energy due to particle collisions. For a monatomic gas, $K_{eh}$ can be described by:

$$K_{eh} = \sum_{s \neq e} \tfrac{3}{2} k_B \frac{2 m_e m_s}{(m_s + m_e)} \left(\frac{8 k_B T_e}{\pi m_e}\right)^{\frac{1}{2}} n_s \sigma_{es}, \tag{22}$$

where $\sigma_{es}$ is the collision cross-section between electrons and the heavy-species $s$, calculated using the Coulomb collision cross-section for electron – ion collisions [39] and the data in [42] for the electron – neutral collision. Figure 3 depicts $S_r$, $S_{eh}$, and $K_{eh}$ as function of $T_e$ and $\theta$ for $p = 1$ [atm]. It can be observed that these terms vary by more than 10 orders of magnitude and that the term $S_{eh}$ dominates over $S_r$ for the presented temperature range.

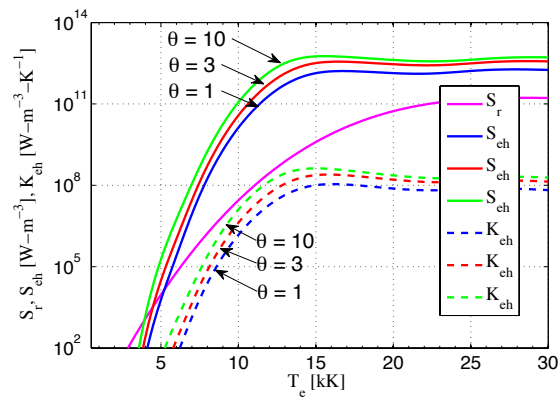

**Figure 3.** Volumetric radiative losses ($S_r$), electron – heavy-species energy exchange source ($S_{eh}$) term and coefficient ($K_{eh}$) for an argon plasma at $p = 1$ [atm] as function of $T_e$ and $\theta$.

## 3. Numerical Model

*3.1. Variational multiscale finite element method*

Equation 11, complemented with appropriate initial and boundary conditions specified over the spatial domain Ω with boundary Γ, give the so-called *strong* form of the arc discharge problem. The *weak*, or Variational, form of the problem is often more suitable for its solution with numerical methods that handle unstructured discretizations naturally, such as the Finite Element Method (FEM), and is given by:

$$(\mathbf{W}, \mathcal{R}(\mathbf{Y}))_\Omega = \mathbf{0}, \qquad (23)$$

where $(\mathbf{A},\mathbf{B})_\Omega \equiv \int_\Omega \mathbf{A}\cdot\mathbf{B}\,d\Omega$ is a bilinear form of **A** and **B**, and **W** is the test function, i.e., any function that belongs to the same mathematical space of **Y** (e.g., space of continuous and bounded functions over Ω).

It is well known that direct numerical solutions of Eq. 11 or Eq. 23 are posed with diverse spurious behavior (e.g., oscillations, instability, and divergence) when the problem is deemed multiscale (i.e., when different terms in Eq. 11 dominate over different parts of Ω; this behavior leads to boundary layers, shocks, chemical fronts, etc.). There are numerous approaches to alleviate these deficiencies, such as Upwinding methods, flux limiters, and Riemann solvers for advection-dominated problems, and stabilized methods and adaptive grid refinement for general TADR problems.

We approach the solution of the TADR arc discharge problem given by Eq. 23 using the Variational Multiscale (VMS) FEM [44], which has been proven very successful in the solution of diverse transport problems, such as incompressible, compressible, reactive, laminar and turbulent flows [44-46], electron – hole transport in semiconductors [48], magnetohydrodynamics [47], and fully ionized plasmas [49, 50]. Initial work in the application of VMS methods for LTE and NLTE thermal plasma flows is reported in [31, 51-53]. The VMS framework is also ideally suited for the modeling of turbulent flows with the same rationality of Large Eddy Simulation (LES) techniques (i.e., solution of the large- and modeling of the small-scales), but with the added advantages of a consistent and complete coarse-grained description of the flow, which is not the case for most traditional LES techniques [46].

The method consists in dividing the solution field **Y** into its large-scale $\bar{\mathbf{Y}}$ and small-scale **Y'** components, i.e.,

$$\mathbf{Y} = \bar{\mathbf{Y}} + \mathbf{Y}', \qquad (24)$$

where the large-scales are solved by the computational discretization and the small-scales, which cannot be described by the discretization, are modeled. Applying the scale decomposition to **Y** and **W**, and hence effectively diving the problem between large- and small-scale sub-problems, and using adjoint duality, Eq. 23 can be expressed in terms of the large- and small-scale terms as:

$$\underbrace{(\bar{\mathbf{W}}, \mathcal{R}(\bar{\mathbf{Y}}))_\Omega}_{\text{large scales}} + \underbrace{(-\mathcal{L}^*\bar{\mathbf{W}}, \mathbf{Y}')_\Omega}_{\text{small scales}} = \mathbf{0}, \tag{25}$$

where $\mathcal{L}$ represents the TADR transport operator, i.e.,

$$\mathcal{L} = \mathbf{A_0}\partial_t + (\mathbf{A}_i\partial_i) - \partial_i(\mathbf{K}_{ij}\partial_j) - \mathbf{S_1}, \tag{26}$$

and $^*$ is the adjoint operator. The small scales are given by the residual-based approximation:

$$\mathbf{Y}' = -\boldsymbol{\tau}\mathcal{R}(\mathbf{Y}), \tag{27}$$

where,

$$\boldsymbol{\tau} \approx \mathcal{L}^{-1}, \tag{28}$$

is an operator that encloses the level of approximation (e.g., if the equal sign is used in Eq. 28, then Eq. 25 provides an *exact* representation of the problem). To obtain a computationally feasible, yet approximate, solution to Eq. 25, an algebraic approximation of $\boldsymbol{\tau}$ is used, in which $\boldsymbol{\tau}$ is described as a matrix of size (*number of unknowns*)$^2$ [44, 46].

Using Eqs. 27 and 26 in Eq. 25, and explicitly separating the temporal discretization from the Finite Element spatial discretization, the following discrete counterpart to Eq. 25 is obtained:

$$\begin{aligned}
&\mathbf{R}(\mathbf{Y}_h, \dot{\mathbf{Y}}_h) = \\
&\underbrace{(\mathbf{N}, \mathbf{A_0}\dot{\mathbf{Y}}_h + \mathbf{A}_i\partial_i\mathbf{Y}_h - \mathbf{S_1}\mathbf{Y}_h - \mathbf{S_0})_{\Omega_h} + (\partial_i\mathbf{N}, \mathbf{K}_{ij}\partial_j\mathbf{Y}_h)_{\Omega_h} - (\mathbf{N}, \mathbf{n}_i\mathbf{K}_{ij}\partial_j\mathbf{Y}_h)_{\Gamma_h}}_{\text{large scales}} + \\
&\underbrace{(\mathbf{A}_i^\mathsf{T}\partial_i\mathbf{N} + \mathbf{S_1}^\mathsf{T}\mathbf{N}, \boldsymbol{\tau}(\mathbf{A_0}\dot{\mathbf{Y}}_h + \mathbf{A}_i\partial_i\mathbf{Y}_h - \mathbf{S_1}\mathbf{Y}_h - \mathbf{S_0}))_{\Omega'_h}}_{\text{small scales}} + \\
&\underbrace{(\partial_i\mathbf{N}, \mathbf{K}_{ij}^{DC}\partial_j\mathbf{Y}_h)_{\Omega_h}}_{\text{discontinuity captuing}} = \mathbf{0}
\end{aligned} \tag{29}$$

where **R** is the discrete counterpart to $\mathcal{R}$, $\mathbf{Y}_h$ is the discrete representation of **Y**, and $\dot{\mathbf{Y}}_h$ its temporal derivative, **N** is the multi-linear (i.e., second-order-accurate) Finite Element basis function (e.g., see [54]), $\Omega_h$ and $\Gamma_h$ represent the discrete spatial domain and its boundary, respectively, **n** is the outer normal to the boundary, and $\Omega'_h$ is a subset of $\Omega_h$ adequate for the description of the small-scales. The third term in the large-scales component represents the imposition of boundary conditions over $\Gamma$. The discontinuity capturing term has been added to increase the robustness of the solution process in regions with large gradients [38]. To maintain the consistency of the formulation, the discontinuity capturing diffusivity matrix $\mathbf{K}_{ij}^{DC}$ is proportional to $\mathcal{R}(\mathbf{Y})$. The small-scales term uses the facts that the spatial discretization is constant in time (therefore, $\partial_t\mathbf{N} = \mathbf{0}$) and that **N** is multi-linear (hence, $\partial_i\partial_i\mathbf{N} = \mathbf{0}$); furthermore, the diffusive part of the residual has not been re-constructed, which is an adequate approximation for the second-order accuracy of the formulation.

Equation 29 represents the VMS-FEM counterpart of Eq 11. Due to the singular nature of $\mathbf{A_0}$ (e.g., see Appendix A), $\mathbf{R}$ represents, in general, a very large differential-algebraic nonlinear system of equations.

*3.2. Solution approach*

To obtain second-order accuracy of the overall formulation, a second-order time-stepper approach is needed for the solution of the differential-algebraic Eq. 29. The solution of Eq. 29 is pursued using the second-order generalized-alpha predictor multi-corrector time-stepper method [55].

Denoting as $n$ the time interval of the current solution (i.e., solution vectors $\mathbf{Y}_h$ and $\dot{\mathbf{Y}}_h$ at time $t_n$), the solution at the next time interval $n + 1$ using the alpha method consists in simultaneously solving the following system of four equations:

$$\mathbf{R}(\mathbf{Y}_{n+\alpha_f}, \dot{\mathbf{Y}}_{n+\alpha_m}) = \mathbf{0}, \tag{30}$$

$$\mathbf{Y}_{n+\alpha_f} = \alpha_f \mathbf{Y}_{n+1} + (1-\alpha_f)\mathbf{Y}_n, \tag{31}$$

$$\dot{\mathbf{Y}}_{n+\alpha_m} = \alpha_m \dot{\mathbf{Y}}_{n+1} + (1-\alpha_m)\dot{\mathbf{Y}}_n, \tag{32}$$

$$\frac{\mathbf{Y}_{n+1} + \mathbf{Y}_n}{\Delta t} = \alpha_g \dot{\mathbf{Y}}_{n+1} + (1-\alpha_g)\dot{\mathbf{Y}}_n, \tag{33}$$

where $\Delta t$ represents the time step size, and $\alpha_f$, $\alpha_m$, and $\alpha_g$ are parameters function of the single user-specified parameter $\alpha \in [0,1]$, the algorithmic parameter of the method that controls high-frequency dissipation. The use of the subscripts $n + \alpha_f$ and $n + \alpha_m$ denote that the solution corresponds to the discrete instants $t_n + \alpha_f \Delta t$ and $t_n + \alpha_g \Delta t$, respectively, and the subscript $h$ has been removed from the vectors $\mathbf{Y}$ and $\dot{\mathbf{Y}}$ to simplify the notation.

Given that the dependence of $\dot{\mathbf{Y}}_{n+1}$ on $\mathbf{Y}_{n+1}$ is specified by the structure of the alpha method (i.e., Eq. 33), Eq. 30 implies the solution of a nonlinear system for $\mathbf{Y}$. This nonlinear system is solved by an inexact Newton method with line-search globalization, i.e.,

$$\| \mathbf{R}^k + \mathbf{J}^k \Delta \mathbf{Y}^k \| \leq \eta^k \| \mathbf{R}^k \|, \text{ and} \tag{34}$$

$$\mathbf{Y}^{k+1} = \mathbf{Y}^k + \lambda^k \Delta \mathbf{Y}^k, \tag{35}$$

where the super-index $k$ represents the iteration counter (e.g., $\mathbf{Y}^k \equiv \mathbf{Y}^k_{n+1}$), $\mathbf{J} \approx \partial \mathbf{R}/\partial \mathbf{Y}$ is the approximate Jacobian, and $\eta$ and $\lambda$ are the tolerance for the solution of the linear system implied by Eq. 34 and the step length, respectively; both calculated according to [56]. The approximate solution of Eq. 34 is accomplished using the Generalized Minimal Residual method [57] using the block-diagonal section of **J** as preconditioner [58].

Two main aspects of the numerical model need to be emphasized:

(1) The spatial – temporal discretization is second-order-accurate in space and time; therefore, it can be expected that, if the size of the elements used in the discretization is reduced by 50% (e.g., using ~ 1.5 times more discretization nodes), then the overall error of the solution should be reduced by a factor of ~ $(1.5)^2 = 2.25$.

(2) The monolithic treatment of the VMS-FEM formulation and solution process (i.e., simultaneous solution of the fluid and electromagnetic field variables) should prevent the reaching of artificial solutions, which are possible in segregated solution algorithms. This aspect is deemed particularly important when solutions prone to be unstable, such as the development of pattern formation, are sought.

The methods above described have been successfully used to describe arc discharge dynamics, including the arc reattachment processes, and plasma jets (e.g., [31, 21]). Details of the implementation and validation of the solver are presented in a forthcoming publication [59].

## 4. The Free-burning Arc

*4.1 Problem description*

The free-burning arc is established by a constant DC current between a conical cathode aligned vertically on top of a flat horizontal anode in the absence of auxiliary gas flow, magnetic confinement, or any other type of external forcing [28]. The self-constriction of the current density over the cathode surface establishes the so-called cathode jet, which accelerates the plasma towards the anode and establishes a recirculating flow of gas and a corresponding bell-shaped optical emission pattern from the plasma. Despite the axi-symmetry of the problem geometry and constancy of operating conditions, different types of instabilities can develop, not only the plasma – anode instabilities that lead to pattern formation, as studied here, but also fluid dynamic instabilities due to the presence of large property gradients (e.g., at the plasma – cold gas interface, or by the flow ejected radially parallel to the anode). This fact has motivated the use of a time-dependent and three-dimensional NLTE model in the present work. As stated in Section 1, the NLTE model is based on the chemical equilibrium assumption and uses relatively simple boundary conditions over the anode surface, and does not include the modeling of the bulk electrodes, electrode sheath models, or anode material evaporation effects [62-69].

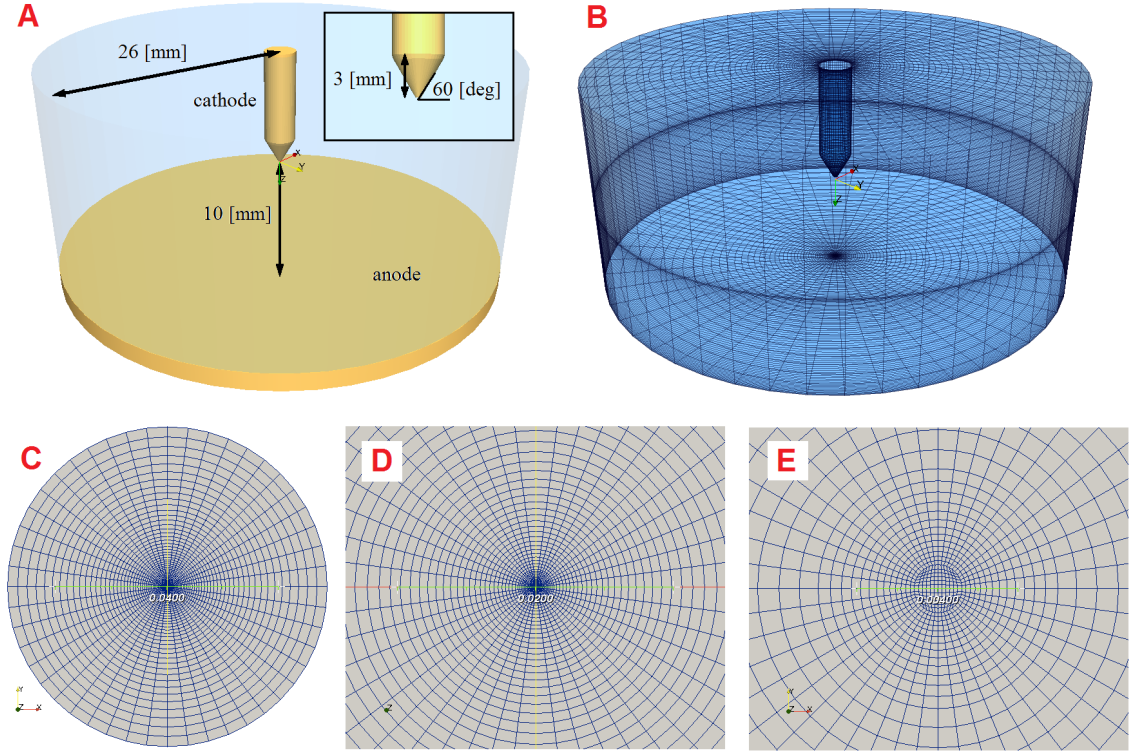

**Figure 4.** Spatial domain of the free-burning arc problem: (A) geometry of the spatial domain for the free-burning arc problem; (B) depiction of the three-dimensional hexahedra finite element *base* mesh; (C) mesh over the anode surface; (D) and (E) different magnifications of the central part of the anode mesh (reference scale at center in units of [m]).

Figure 4 presents the spatial domain $\Omega$ (e.g., 10 [mm] inter-electrode spacing, 60 [deg] conical cathode) and different views of the computational mesh. The boundary of the domain $\Gamma$ is composed of: (1) the *cathode* surface, (2) the *anode* surface, and (3) the surrounding *open flow*. Two discretization grids have been used for the simulations: a *base* mesh for the majority of the analyses (depicted in Fig. 4, Frames B to E), and a *fine* mesh for verification/validation (not shown in Fig. 4). Both meshes are topologically similar. For the *base* mesh, the domain is discretized with ~ $2.5 \; 10^5$ tri-linear hexahedral elements and ~ $2.6 \; 10^5$ nodes (i.e., Eq. 29 implies the simultaneous solution of ~ $2.6 \; 10^6$ nonlinear equations at each time step); the mesh is finer near the cathode tip and stretched towards the open flow boundaries. The minimum and maximum lengths for any element are approximately 0.03 and 1.7 [mm], respectively. For the *fine* mesh, the domain is discretized with ~ 4.0 10$^5$ tri-linear hexahedral elements and ~ 4.1 10$^5$ nodes; therefore, the fine mesh has approximately 2 times the resolution and size of the base mesh. Frames C to E in Fig. 4 show the part of the base mesh covering the anode surface for different levels of magnification. It can be noticed that the center of the mesh is not singular, and moreover, that it is discretized by quasi-equilateral hexahedra. Both characteristics help prevent numerical instabilities in the solution. Furthermore, it is well known that the effect of the spatial resolution on the obtained results is particularly significant near the electrode region in arc discharge problems, e.g., [64, 65].

*4.2 Boundary conditions*

The set of boundary conditions used, consisted with the NLTE model, is listed in Table 5.

**Table 5.** Set of boundary conditions for each variable for the free-burning arc problem.

| Boundary | Variable | | | | | |
|---|---|---|---|---|---|---|
| | $p$ | $\mathbf{u}$ | $T_h$ | $T_e$ | $\phi_p$ | $\mathbf{A}$ |
| Cathode | $\partial_n p = 0$ | $\mathbf{u} = \mathbf{0}$ | $T_h = T_{cath}(z)$ | $\partial_n T_e = 0$ | $-\sigma \partial_n \phi_p = J_{qcath}(r)$ | $\partial_n \mathbf{A} = \mathbf{0}$ |
| Anode | $\partial_n p = 0$ | $\mathbf{u} = \mathbf{0}$ | $-\kappa_h \partial_n T_h = h_w(T_h - T_w)$ | $\partial_n T_e = 0$ | $\phi_p = 0$ | $\partial_n \mathbf{A} = \mathbf{0}$ |
| Open flow | $p = p_\infty$ | $\partial_n \mathbf{u} = \mathbf{0}$ | $T_h = T_\infty$ | $T_e = T_\infty$ | $\partial_n \phi_p = 0$ | $\partial_n \mathbf{A} = \mathbf{0}$ |

In Table 5, $\partial_n \equiv \mathbf{n} \cdot \nabla$, with $\mathbf{n}$ as the outer normal, denotes the derivative normal to the surface; $p_\infty$ is the reference open flow pressure, set equal to the atmospheric pressure (1.01325·10$^5$ [Pa]), and $T_\infty$ = 500 [K] is a reference open flow temperature.

The temperature profile imposed over the cathode surface $T_{cath}$ is given by:

$$T_{cath} = T_{crod} + (T_{ctip} - T_{crod})\exp(-(z/L_{cath})^2), \qquad (36)$$

where $T_{crod}$ and $T_{ctip}$ are the temperatures of the cathode rod and tip, equal to 500 [K] and 3600 [K] respectively, and $L_{cath}$ is a characteristic length set equal to 1.5 [mm].

Heat transfer to the anode is modeled assuming convective heat losses in a water-cooled metal anode using $h_w$ = 10$^5$ [W-m$^{-2}$-K$^{-1}$] as the convective heat transfer coefficient and $T_w$ = 500 [K] as the reference cooling water temperature. This boundary condition is different from that used in [29], but is often adopted in arc plasma flow simulations, e.g., [31, 21, 72].

The current density profile over the cathode $J_{qcath}$ is given by:

$$J_{qcath} = J_{qmax} \exp(-(r/r_{cath})^{n_{cath}}),  \quad (37)$$

where $r = (x^2 + y^2)^{\frac{1}{2}}$ is the radial coordinate, and $J_{qmax}$, $r_{cath}$, and $n_{cath}$ are parameters that control the shape of the current density profile, which has to satisfy the imposition of the total electric current to the system, i.e., $I_{tot} = \int_{S_{cath}} J_{qcath} dS$, where $S_{cath}$ represents the cathode surface. Table 6 shows the values of $J_{qmax}$ and $r_{cath}$ for the values of $I_{tot}$ simulated (i.e., from 100 to 300 [A] in intervals of 25 [A]). A value of $n_{cath} = 4$ has been used for all $J_{qcath}$ profiles, whereas a value of $n_{cath} = 1$ has been traditionally used for sharply conical cathodes or truncated two-dimensional computational domains (e.g., [29, 68, 76]). The value of $J_{qmax}$ at 200 [A] is higher than the value of ~ 1.6 [$10^8$ A-m$^{-2}$] reported by Lowke and collaborators [70] from simulations that included modeling of the electrodes. The dependency of $J_{qmax}$ on $I_{tot}$ has been chosen such that $(J_{qcath}I_{tot})^{\frac{1}{2}}$ varies approximately linearly with $I_{tot}$; this functional dependency produces the expected behavior of the cathode jet [28], as described in the following section.

**Table 6.** Parameters for the specification of the current density profile over the cathode.

|  | $I_{tot}$ [A] | | | | | | | | |
| --- | --- | --- | --- | --- | --- | --- | --- | --- | --- |
|  | 100 | 125 | 150 | 175 | 200 | 225 | 250 | 275 | 300 |
| $J_{qmax}$ [$10^8$ A-m$^{-2}$] | 2.000 | 2.250 | 2.250 | 2.375 | 2.500 | 2.625 | 2.750 | 2.875 | 3.000 |
| $r_{cath}$ [$10^{-4}$ m] | 3.310 | 3.458 | 3.736 | 3.901 | 4.043 | 4.168 | 4.278 | 4.376 | 4.464 |

**5. Simulations of Anode Patterns**

*5.1 Verification and validation*

Figure 5 shows computational results using the base mesh for $I_{tot} = 200$ [A]. Frame A of Fig. 5 depicts the distribution of $T_h$ over the whole domain, where it can be observed that the extent of the domain is large enough to adequately describe open flow boundary conditions. Frame B shows the distribution of $T_h$ in the inter-electrode region, where the numerical results obtained in the present work (*right*) are contrasted with the experimentally measured equilibrium temperature $T$ from Hsu and Pfender [29] (*left*). The modeling results show higher constriction of the arc compared to the experimental measurements in [29]. The higher constriction may be due to two effects: (1) the larger values of $J_{qmax}$ used with respect to those in [29], and (2) the stronger cooling of the anode surface, evidenced by the difference in slopes of the $T$ and $T_h$ iso-contours near the anode. It should be noticed that the simulations by Hsu and Pfender used experimentally measured temperatures near the anode as boundary condition, whereas the convective cooling condition used here (Table 5) is more applicable to general arc discharge simulations. The effect of the stronger cooling of the anode is addressed in greater detail in the next section.

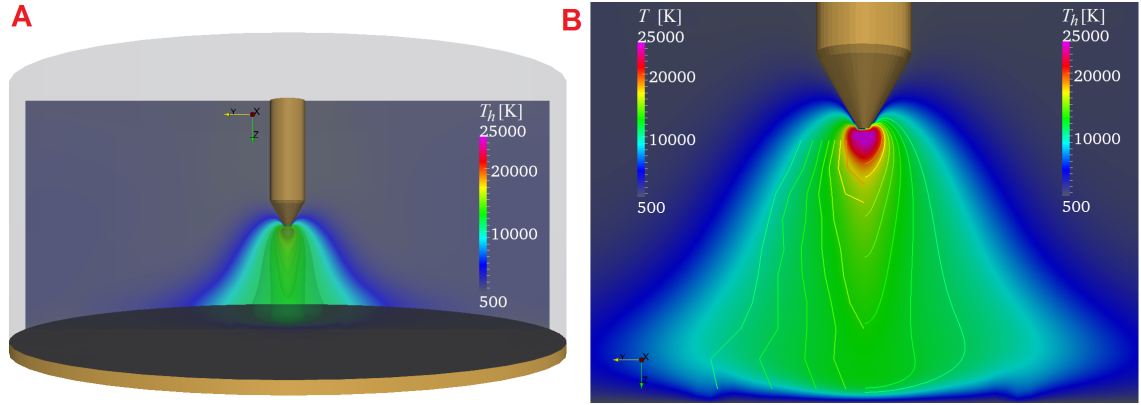

**Figure 5.** Temperature distribution for $I_{tot}$ = 200 [A]: (A) distribution of $T_h$ over the complete spatial domain; and (B) inter-electrodes region: (*left*) iso-contours of experimental temperature measurements from [29] and (*right*) simulation results.

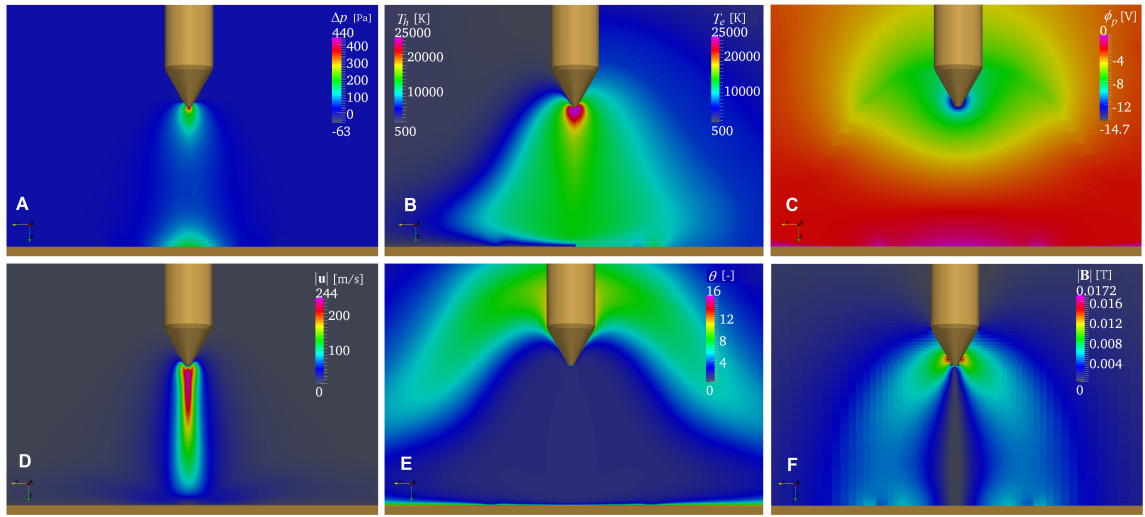

**Figure 6.** Solution fields for the free-burning arc at $I_{tot}$ = 200 [A]: (A) pressure difference $\Delta p = p - p_\infty$; (B) heavy-species and electron temperature, $T_h$ (*left*) and $T_e$ (*right*), respectively; (C) effective vector potential $\phi_p$; (D) magnitude of velocity vector $|\mathbf{u}|$; (E) nonequilibrium parameter $\theta = T_e/T_h$; and (F): magnitude of magnetic field $\mathbf{B} = \nabla \times \mathbf{A}$.

Figure 6 shows the distribution of different fields associated to the solution for $I_{tot} = 200$ [A]. The distribution of pressure difference $\Delta p = p - p_\infty$ in Frame A is consistent with the formation of the cathode jet in Frame D; both results are compatible with previously reported numerical results, e.g., [29, 65, 74]. Frame B contrasts the distributions of $T_h$ and $T_e$, where the more diffuse character of $T_e$, consistent with the boundary conditions used, produces the distribution of the thermodynamic nonequilibirum parameter $\theta$ in Frame E. The high values of $\theta$ in the arc fringes are in consistent with the analysis reported in [60]. Frame C shows the distribution of the effective voltage $\phi_p$ and indicates that the maximum voltage drop is -14.7 [V]; this value is higher than the -13.3 [V] reported in [29, 65], but lower than the -16 [V] reported in [73]. Frame F presents the distribution of the magnitude of magnetic field $\mathbf{B} = \nabla \times \mathbf{A}$, which produces the magnetic pumping leading to the cathode jet. The small perturbations in |**B**| near the anode are due to the occurrence of anode patterns, as described in the next section.

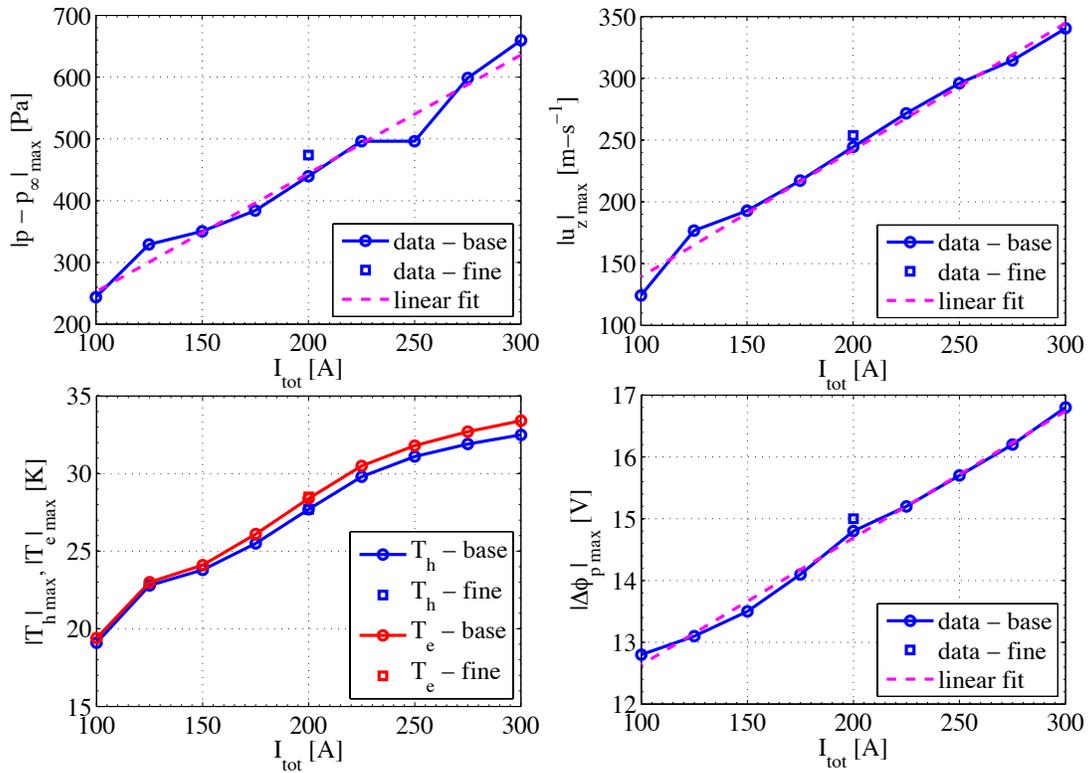

**Figure 7.** Maximum values of solution fields as function of total current $I_{tot}$ from simulation data using the base mesh (*data - base*), the finer mesh (*data - fine*), and the linear fit of the results (*linear fit*): (*top-left*) pressure difference $\Delta p = p - p_\infty$, (*top-right*) velocity component along the $z$

axis $|u_z|$, (*bottom-left*) heavy-species $T_h$ and electron $T_e$ temperature, and (*bottom-right*) total voltage drop $|\Delta\phi_p|=|\phi_p|$.

Figure 7 presents the variation of the maximum values of pressure difference, axial velocity, temperatures, and voltage drop with total current using the base mesh, as well as the results using the fine mesh for $I_{tot}$ = 200 [A]. Frames A, B, and C provide characteristics of the cathode jet (the acceleration of the flow due to the constriction caused by magnetic pinching), whereas frame D is a characteristic of the overall arc. The quasi-linear dependency of $u_{zmax}$ with $I_{tot}$ is consistent with the expected behavior $u_{z\max} \propto (J_{qcath}I_{tot})^{\frac{1}{2}}$ described in [28], given the variation of $J_{qmax}$ on $I_{tot}$ listed in Table 6. The quadratic dependence of the maximum temperatures and the linear dependence of voltage drop with $I_{tot}$ are in agreement with the numerical results from Lowke *et al* [70] and the experiments by Haidar [71]. The higher maximum temperatures in the present work are caused by the higher values of $J_{qmax}$ compared to those in [70]. It can be noticed that the deviation between the maximum heavy-species and electron temperatures increases with total current. The relatively difference between the results using the base and the fine meshes, even though the latter is nearly 2 times finer, is very small, being less than 8% for the $p$, less than 4% for $u_z$, and in the order of 1% for the other fields.

*5.2 Self-organization of anode spot patterns*

The occurrence of anode patterns is visualized in Fig. 8 by the distribution of $T_h$ in the *x-y* plane at 0.2 [mm] away from the anode along the *z*-axis. Frame A in Fig. 8 depicts a three-dimensional view of the arc given by iso-surfaces of $T_h$, whereas Frame B shows a detail of the region near the anode and the plane from which the patterns are extracted. Frame C shows the set of anode patterns for the 9 values of $I_{tot}$ simulated, from 300 [A] (Sub-frame 1) to 100 [A] (Sub-frame 9) in steps of 25 [A]. It can be observed that, for higher values of $I_{tot}$, there is a well-defined major attachment spot at the center of the anode surrounded by small attachment spots. As $I_{tot}$ decreases, the center attachment weakens, which leads to the subsequent formation of additional spots in its place. These results indicate that the anode spots originate at the fringes of the arc, corroborating the conclusions reported in [17] from a stability analysis and experimental observations. Numerical solutions for $I_{tot}$ less than 100 [A] were nonstationary, whereas solutions $I_{tot}$ greater than 300 [A] display high temperatures that limit the validity of the four-species Ar plasma model (e.g., $Ar^{+++}$ species would need to be considered), and therefore are not presented here. The lack of perfect symmetry of the patterns may be due to the use of an iterative linear solver (i.e., Eq.

34) for the solution of Eq. 29, which, although essential for the solution of large systems of equations (2.6 $10^6$ unknowns in this case), may compromise the accuracy of the final solution.

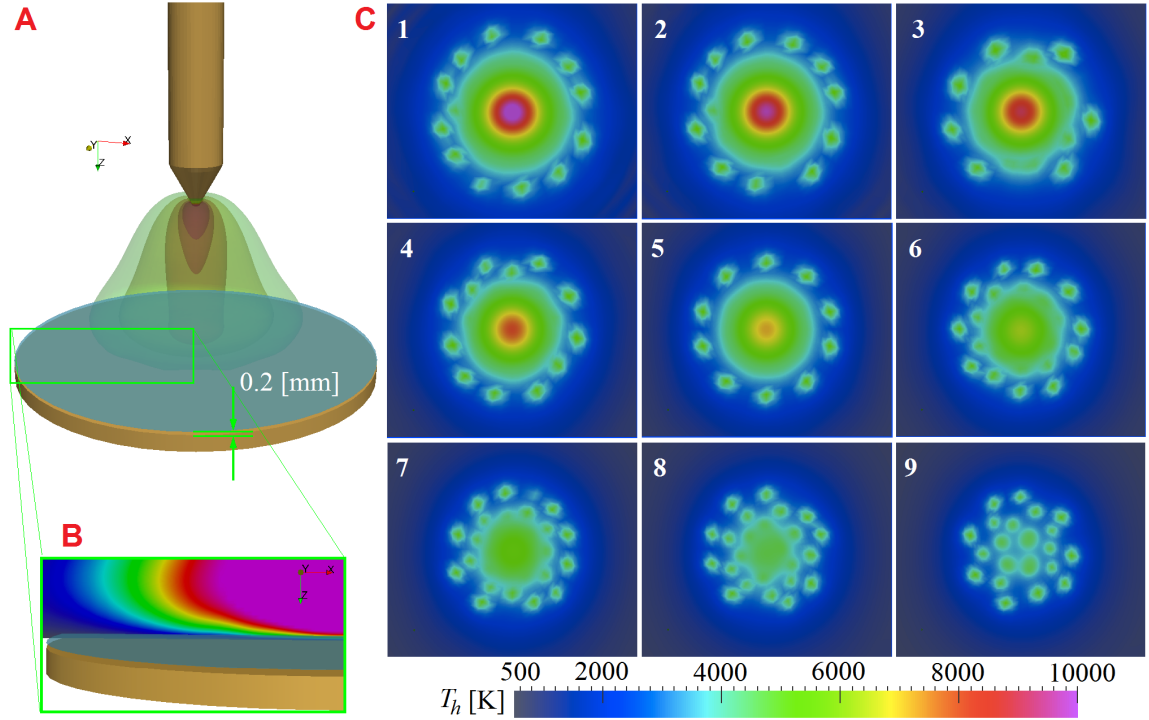

**Figure 8.** Self-organization of anode patterns: (A) $T_h$ iso-contours at 8, 10, 12, 14, and 16 [kK] and cutting plane at 0.2 [mm] above the anode used to visualize the anode spots; (B) $T_h$ distribution near the anode for $I_{tot}$ = 200 [A]; and (C) anode patterns for total current $I_{tot}$ from 300 to 100 [A] in intervals of 25 [A] (i.e., $I_{tot}$ = 300 [A] for frame 1, $I_{tot}$ = 275 [A] for frame 2, etc.).

The relative strength of the spots, given by their average value of $T_h$, is nearly constant in all cases and around 5000 [K]. In contrast, the distribution of $T_h$ at the anode surface (i.e., 0.0 [mm] from the anode) is dominated by the $T_h$ boundary condition (Table 5). For $I_{tot}$ = 200 [A], the average maximum temperature at the center of the anode surface is approximately 750 [K] and slightly higher in the surrounding spots. This value is significantly lower that the ~ 900 [K] reported in [63] using detailed LTE steady-state axi-symmetric simulations that included the modeling of heat transfer through the electrodes (not showing anode attachment patterns). These lower temperatures indicate that the simulations presented here model stronger cooling of the anode surface, which may emphasize the formation of anode patterns.

Figure 9 shows the experimental results reported by Yand and Heberlein in [17] of the planetary distribution of anode burn patterns caused by constricted anode attachments in a forced transferred arc operating at 100 [A] (Frame A), together with the numerical results for 200 [A] using the base and fine meshes (Frames B and C, respectively). Despite the higher current and lack of flow forcing, the numerical results capture the planetary distribution of anode patterns with a stronger, dominant, attachment spot at the center of the anode.

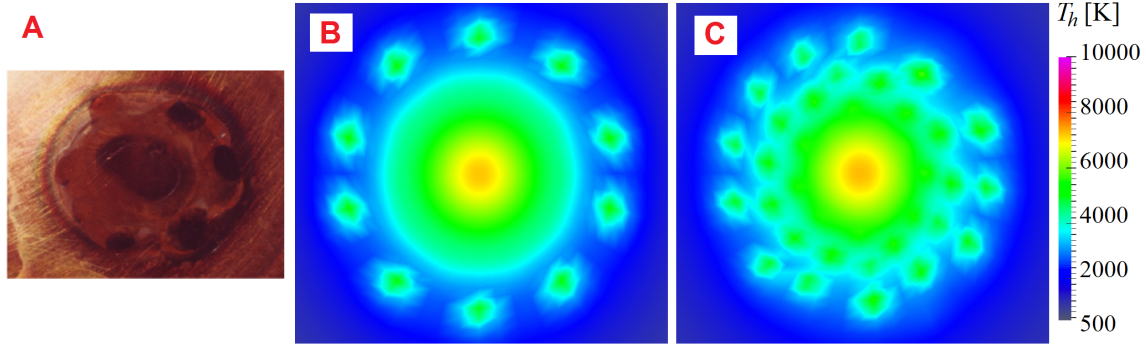

**Figure 9.** Validation of anode patterns in the free-burning arc: (A) results from the experiments by Yand and Heberlein (forced arc, 100 [A] and flow rates from 5 to 15 [slpm]) [17]; (B) and (C) simulation results of $T_h$ at 0.2 [mm] above the anode for $I_{tot}$ = 200 [A] suing the base and fine meshes, respectively.

The results in Frames B and C of Fig. 9 indicate clear differences in the obtained anode spots using different degrees of spatial discretization (i.e., base and fine meshes). The marked difference in anode patterns is somewhat surprising considering that overall solution quantities vary by less than 10% between meshes (see Section 5.1 and Fig. 7). The patterns differ in the number, size, and location of the anode spots, but not in the average value of $T_h$ in them or the overall $T_h$ distribution. This fact may be indicative that spots are strongly correlated with the spatial discretization (e.g., that spots may form at discretization nodes) and that thermodynamic nonequilibrium plays a major role in the overall current transfer to the anode (explaining the lack of difference in $T_h$ distribution for the base and fine meshes). Both conjectures are addressed by the results in Fig. 10 below.

The authors in [17] stated that the formation of constricted spots is in part due to the evaporation of anode material; an effect that is not accounted for in the numerical simulations presented here. The effect of electrode material evaporation on the arc has been reported [61], where numerical simulations indicate that the addition of metal vapor from the cathode increases the electrical conductivity of the plasma, increasing its core temperature and constricting the size of its attachment to the anode. Nevertheless, the fact that the spot patterns can be captured with the present model indicate that other effect may also drive the formation of anode patterns, particularly, the degree of thermodynamic nonequilibrium. Figure 10 shows the distribution of the nonequilibrium parameter $\theta$ corresponding to the $T_h$ plots in Fig. 8.

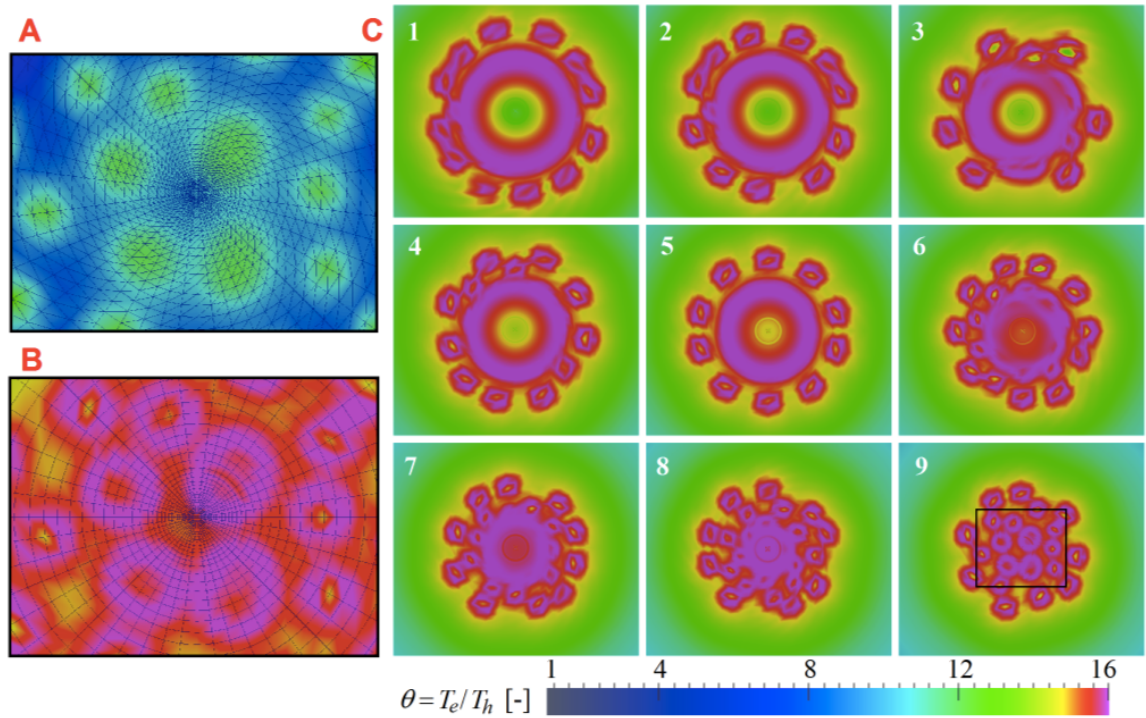

**Figure 10.** Nonequilibrium in anode patterns: (A) detail of the distribution of $T_h$ and (B) of $\theta = T_e/T_h$ at 0.2 [mm] above the anode surface for $I_{tot}$ = 100 [A] showing that the resolution of each spot encompasses several grid points for the base mesh; and (C) distribution of $\theta$ at 0.2 [mm] above the anode surface for $I_{tot}$ from 300 to 100 [A] in intervals of 25 [A]; the box in frame 9 depicts the extent of domain in Frames A and B.

Frame A of Fig. 10 shows a detail of the distribution of $T_h$ near the center of the anode for $I_{tot}$ = 100 [A] (i.e., region indicated by a rectangle in Frame C - Sub-Frame 9), the value of $I_{tot}$ that displays the highest number of spots; whereas the results in Frame B show the corresponding distribution of $\theta$. The results in Frames A and B are superimposed to the computational grid used for the spatial discretization (i.e., base mesh), showing that the anode spots encompass several discretization nodes. The same is true for the results using the fine mesh. Therefore, the occurrence of anode spots is not strongly correlated to the distribution of grid nodes. The results in Frame C indicate that a thermodynamic nonequilibrium "ring" surrounds each anode spot, particularly the main spot at the center of the anode for the higher values of current. The degree of nonequilbrium remains high in the regions between spots, effectively increasing the area for current transfer. The high degree of thermodynamics nonequilibrium is likely due to the need to maintain continuity of current transfer in spite of the intense cooling of the anode surface.

## 6. Conclusions

Time-dependent three-dimensional thermodynamic nonequilibrium simulations reveal the spontaneous formation of self-organized patterns of anode attachment spots in the free-burning arc. The number of spots, their size, and distribution within the pattern depend on the applied total current and on the resolution of the spatial discretization. The results indicate that: (1) the formation of anode spots can be captured by a thermodynamic nonequilibrium, three-dimensional and time-dependent arc discharge model, (2) the occurrence of spots does not depend on accounting for the effect of metal vapors, chemical nonequilibrium, sheath models, or complex boundary conditions, (3) the location and number of spots depends on the resolution of the computational grid. The sensibility of the solution to the spatial discretization stresses the numerical (e.g., second-order) and computational (e.g., time-dependent, three-dimensional, and high spatial resolution) requirements for comprehensive arc discharge simulations. The limited symmetry of some of the obtained anode patterns may be due to the use of an iterative (approximate) linear solver; it can be expected that the use of parallel direct solvers may produce higher quality solutions depicting a higher degree of symmetry. The simulation results show that the anode spots originate at the fringes of the arc – anode attachment, corroborating the analysis reported in [17], and that the heavy-species – electron energy equilibration, in addition to thermal instability, has a dominant role in the formation of anode spots in arc discharges.

## Appendix A. TADR matrices

The transport matrices $\mathbf{A_0}$, $\mathbf{A}_i$, $\mathbf{K}_{ij}$, $\mathbf{S_1}$ and vector $\mathbf{S_0}$ define the NLTE plasma flow model. For three-dimensional problems, the indexes $i, j$ represent the Cartesian coordinates $\{x, y, z\}$, and the vector of primitive variables is given by:

$$\mathbf{Y} = [\, p \quad u_x \quad u_y \quad u_z \quad T_h \quad T_e \quad \phi_p \quad A_x \quad A_y \quad A_z \,]^\mathrm{T}. \tag{A.1}$$

The coefficients below are used to simplify the explicit notation of the matrices:

$$\rho^p = \frac{\partial \rho}{\partial p}, \; \rho^h = \frac{\partial \rho}{\partial T_h}, \; \rho^e = \frac{\partial \rho}{\partial T_e}, \tag{A.2}$$

$$C_h^p = \rho \frac{\partial h_h}{\partial p}, \quad C_h^h = \rho \frac{\partial h_h}{\partial T_h}, \quad C_h^e = \rho \frac{\partial h_h}{\partial T_e}, \tag{A.3}$$

$$C_e^p = \rho \frac{\partial h_e}{\partial p}, \quad C_e^h = \rho \frac{\partial h_e}{\partial T_h}, \quad C_e^e = \rho \frac{\partial h_e}{\partial T_e}, \text{ and} \tag{A.4}$$

$$\sigma_\mu = \mu_0 \sigma. \tag{A.5}$$

Using the above coefficients, the transport matrices of the TADR system are:

$$\mathbf{A_0} = \begin{bmatrix} \rho^p & 0 & 0 & 0 & \rho^h & \rho^e & 0 & 0 & 0 & 0 \\ 0 & \rho & 0 & 0 & 0 & 0 & 0 & 0 & 0 & 0 \\ 0 & 0 & \rho & 0 & 0 & 0 & 0 & 0 & 0 & 0 \\ 0 & 0 & 0 & \rho & 0 & 0 & 0 & 0 & 0 & 0 \\ C_h^p & 0 & 0 & 0 & C_h^h & C_h^e & 0 & 0 & 0 & 0 \\ C_e^p & 0 & 0 & 0 & C_e^h & C_e^e & 0 & 0 & 0 & 0 \\ 0 & 0 & 0 & 0 & 0 & 0 & 0 & 0 & 0 & 0 \\ 0 & 0 & 0 & 0 & 0 & 0 & 0 & \sigma_\mu & 0 & 0 \\ 0 & 0 & 0 & 0 & 0 & 0 & 0 & 0 & \sigma_\mu & 0 \\ 0 & 0 & 0 & 0 & 0 & 0 & 0 & 0 & 0 & \sigma_\mu \end{bmatrix}, \tag{A.6}$$

$$\mathbf{A_x} = \begin{bmatrix} \rho^p u_x & \rho & 0 & 0 & \rho^h u_x & \rho^e u_x & 0 & 0 & 0 & 0 \\ 1 & \rho u_x & 0 & 0 & 0 & 0 & 0 & 0 & 0 & 0 \\ 0 & 0 & \rho u_x & 0 & 0 & 0 & 0 & 0 & 0 & 0 \\ 0 & 0 & 0 & \rho u_x & 0 & 0 & 0 & 0 & 0 & 0 \\ C_h^p u_x & 0 & 0 & 0 & C_h^h u_x & C_h^e u_x & 0 & 0 & 0 & 0 \\ C_e^p u_x & 0 & 0 & 0 & C_e^h u_x & C_e^e u_x & 0 & 0 & 0 & 0 \\ 0 & 0 & 0 & 0 & 0 & 0 & 0 & 0 & 0 & 0 \\ 0 & 0 & 0 & 0 & 0 & 0 & \sigma_\mu & 0 & -\sigma_\mu u_y & -\sigma_\mu u_z \\ 0 & 0 & 0 & 0 & 0 & 0 & 0 & 0 & \sigma_\mu u_x & 0 \\ 0 & 0 & 0 & 0 & 0 & 0 & 0 & 0 & 0 & \sigma_\mu u_x \end{bmatrix}, \tag{A.7}$$

$$\mathbf{A_y} = \begin{bmatrix} \rho^p u_y & 0 & \rho & 0 & \rho^h u_y & \rho^e u_y & 0 & 0 & 0 & 0 \\ 0 & \rho u_y & 0 & 0 & 0 & 0 & 0 & 0 & 0 & 0 \\ 1 & 0 & \rho u_y & 0 & 0 & 0 & 0 & 0 & 0 & 0 \\ 0 & 0 & 0 & \rho u_y & 0 & 0 & 0 & 0 & 0 & 0 \\ C_h^p u_y & 0 & 0 & 0 & C_h^h u_y & C_h^e u_y & 0 & 0 & 0 & 0 \\ C_e^p u_y & 0 & 0 & 0 & C_e^h u_y & C_e^e u_y & 0 & 0 & 0 & 0 \\ 0 & 0 & 0 & 0 & 0 & 0 & 0 & 0 & 0 & 0 \\ 0 & 0 & 0 & 0 & 0 & 0 & 0 & \sigma_\mu u_y & 0 & 0 \\ 0 & 0 & 0 & 0 & 0 & 0 & \sigma_\mu & -\sigma_\mu u_x & 0 & -\sigma_\mu u_z \\ 0 & 0 & 0 & 0 & 0 & 0 & 0 & 0 & 0 & \sigma_\mu u_y \end{bmatrix}, \tag{A.8}$$

$$\mathbf{A}_z = \begin{bmatrix} \rho^p u_z & 0 & 0 & \rho & \rho^h u_z & \rho^e u_z & 0 & 0 & 0 & 0 \\ 0 & \rho u_z & 0 & 0 & 0 & 0 & 0 & 0 & 0 & 0 \\ 0 & 0 & \rho u_z & 0 & 0 & 0 & 0 & 0 & 0 & 0 \\ 1 & 0 & 0 & \rho u_z & 0 & 0 & 0 & 0 & 0 & 0 \\ C_h^p u_z & 0 & 0 & 0 & C_h^h u_z & C_h^e u_z & 0 & 0 & 0 & 0 \\ C_e^p u_z & 0 & 0 & 0 & C_e^h u_z & C_e^e u_z & 0 & 0 & 0 & 0 \\ 0 & 0 & 0 & 0 & 0 & 0 & 0 & 0 & 0 & 0 \\ 0 & 0 & 0 & 0 & 0 & 0 & 0 & \sigma_\mu u_z & 0 & 0 \\ 0 & 0 & 0 & 0 & 0 & 0 & 0 & 0 & \sigma_\mu u_z & 0 \\ 0 & 0 & 0 & 0 & 0 & 0 & \sigma_\mu & -\sigma_\mu u_x & -\sigma_\mu u_y & 0 \end{bmatrix}, \quad (A.9)$$

$$\mathbf{K}_{xx} = \begin{bmatrix} 0 & 0 & 0 & 0 & 0 & 0 & 0 & 0 & 0 & 0 \\ 0 & \tfrac{4}{3}\mu & 0 & 0 & 0 & 0 & 0 & 0 & 0 & 0 \\ 0 & 0 & \mu & 0 & 0 & 0 & 0 & 0 & 0 & 0 \\ 0 & 0 & 0 & \mu & 0 & 0 & 0 & 0 & 0 & 0 \\ 0 & 0 & 0 & 0 & \kappa_{hr} & 0 & 0 & 0 & 0 & 0 \\ 0 & 0 & 0 & 0 & 0 & \kappa_e & 0 & 0 & 0 & 0 \\ 0 & 0 & 0 & 0 & 0 & 0 & \sigma & 0 & -\sigma u_y & -\sigma u_z \\ 0 & 0 & 0 & 0 & 0 & 0 & 0 & 1 & 0 & 0 \\ 0 & 0 & 0 & 0 & 0 & 0 & 0 & 0 & 1 & 0 \\ 0 & 0 & 0 & 0 & 0 & 0 & 0 & 0 & 0 & 1 \end{bmatrix}, \quad (A.10)$$

$$\mathbf{K}_{xy} = \begin{bmatrix} 0 & 0 & 0 & 0 & 0 & 0 & 0 & 0 & 0 & 0 \\ 0 & 0 & -\tfrac{2}{3}\mu & 0 & 0 & 0 & 0 & 0 & 0 & 0 \\ 0 & \mu & 0 & 0 & 0 & 0 & 0 & 0 & 0 & 0 \\ 0 & 0 & 0 & 0 & 0 & 0 & 0 & 0 & 0 & 0 \\ 0 & 0 & 0 & 0 & 0 & 0 & 0 & 0 & 0 & 0 \\ 0 & 0 & 0 & 0 & 0 & 0 & 0 & 0 & 0 & 0 \\ 0 & 0 & 0 & 0 & 0 & 0 & 0 & \sigma u_y & 0 & 0 \\ 0 & 0 & 0 & 0 & 0 & 0 & 0 & 0 & 0 & 0 \\ 0 & 0 & 0 & 0 & 0 & 0 & 0 & 0 & 0 & 0 \\ 0 & 0 & 0 & 0 & 0 & 0 & 0 & 0 & 0 & 0 \end{bmatrix}, \quad (A.11)$$

$$\mathbf{K}_{xz} = \begin{bmatrix} 0 & 0 & 0 & 0 & 0 & 0 & 0 & 0 & 0 & 0 \\ 0 & 0 & 0 & -\tfrac{2}{3}\mu & 0 & 0 & 0 & 0 & 0 & 0 \\ 0 & 0 & 0 & 0 & 0 & 0 & 0 & 0 & 0 & 0 \\ 0 & \mu & 0 & 0 & 0 & 0 & 0 & 0 & 0 & 0 \\ 0 & 0 & 0 & 0 & 0 & 0 & 0 & 0 & 0 & 0 \\ 0 & 0 & 0 & 0 & 0 & 0 & 0 & 0 & 0 & 0 \\ 0 & 0 & 0 & 0 & 0 & 0 & 0 & \sigma u_z & 0 & 0 \\ 0 & 0 & 0 & 0 & 0 & 0 & 0 & 0 & 0 & 0 \\ 0 & 0 & 0 & 0 & 0 & 0 & 0 & 0 & 0 & 0 \\ 0 & 0 & 0 & 0 & 0 & 0 & 0 & 0 & 0 & 0 \end{bmatrix}, \quad (A.12)$$

$$\mathbf{K}_{yx} = \begin{bmatrix} 0 & 0 & 0 & 0 & 0 & 0 & 0 & 0 & 0 & 0 \\ 0 & 0 & \mu & 0 & 0 & 0 & 0 & 0 & 0 & 0 \\ 0 & -\tfrac{2}{3}\mu & 0 & 0 & 0 & 0 & 0 & 0 & 0 & 0 \\ 0 & 0 & 0 & 0 & 0 & 0 & 0 & 0 & 0 & 0 \\ 0 & 0 & 0 & 0 & 0 & 0 & 0 & 0 & 0 & 0 \\ 0 & 0 & 0 & 0 & 0 & 0 & 0 & 0 & 0 & 0 \\ 0 & 0 & 0 & 0 & 0 & 0 & 0 & 0 & \sigma u_x & 0 \\ 0 & 0 & 0 & 0 & 0 & 0 & 0 & 0 & 0 & 0 \\ 0 & 0 & 0 & 0 & 0 & 0 & 0 & 0 & 0 & 0 \\ 0 & 0 & 0 & 0 & 0 & 0 & 0 & 0 & 0 & 0 \end{bmatrix}, \quad (A.13)$$

$$\mathbf{K}_{yy} = \begin{bmatrix} 0 & 0 & 0 & 0 & 0 & 0 & 0 & 0 & 0 & 0 \\ 0 & \mu & 0 & 0 & 0 & 0 & 0 & 0 & 0 & 0 \\ 0 & 0 & \tfrac{4}{3}\mu & 0 & 0 & 0 & 0 & 0 & 0 & 0 \\ 0 & 0 & 0 & \mu & 0 & 0 & 0 & 0 & 0 & 0 \\ 0 & 0 & 0 & 0 & \kappa_{hr} & 0 & 0 & 0 & 0 & 0 \\ 0 & 0 & 0 & 0 & 0 & \kappa_e & 0 & 0 & 0 & 0 \\ 0 & 0 & 0 & 0 & 0 & 0 & \sigma & -\sigma u_x & 0 & -\sigma u_z \\ 0 & 0 & 0 & 0 & 0 & 0 & 0 & 1 & 0 & 0 \\ 0 & 0 & 0 & 0 & 0 & 0 & 0 & 0 & 1 & 0 \\ 0 & 0 & 0 & 0 & 0 & 0 & 0 & 0 & 0 & 1 \end{bmatrix}, \quad (A.14)$$

$$\mathbf{K}_{yz} = \begin{bmatrix} 0 & 0 & 0 & 0 & 0 & 0 & 0 & 0 & 0 & 0 \\ 0 & 0 & 0 & 0 & 0 & 0 & 0 & 0 & 0 & 0 \\ 0 & 0 & 0 & -\tfrac{2}{3}\mu & 0 & 0 & 0 & 0 & 0 & 0 \\ 0 & 0 & \mu & 0 & 0 & 0 & 0 & 0 & 0 & 0 \\ 0 & 0 & 0 & 0 & 0 & 0 & 0 & 0 & 0 & 0 \\ 0 & 0 & 0 & 0 & 0 & 0 & 0 & 0 & 0 & 0 \\ 0 & 0 & 0 & 0 & 0 & 0 & 0 & 0 & \sigma u_z & 0 \\ 0 & 0 & 0 & 0 & 0 & 0 & 0 & 0 & 0 & 0 \\ 0 & 0 & 0 & 0 & 0 & 0 & 0 & 0 & 0 & 0 \\ 0 & 0 & 0 & 0 & 0 & 0 & 0 & 0 & 0 & 0 \end{bmatrix}, \quad (A.15)$$

$$\mathbf{K}_{zx} = \begin{bmatrix} 0 & 0 & 0 & 0 & 0 & 0 & 0 & 0 & 0 & 0 \\ 0 & 0 & 0 & \mu & 0 & 0 & 0 & 0 & 0 & 0 \\ 0 & 0 & 0 & 0 & 0 & 0 & 0 & 0 & 0 & 0 \\ 0 & -\tfrac{2}{3}\mu & 0 & 0 & 0 & 0 & 0 & 0 & 0 & 0 \\ 0 & 0 & 0 & 0 & 0 & 0 & 0 & 0 & 0 & 0 \\ 0 & 0 & 0 & 0 & 0 & 0 & 0 & 0 & 0 & 0 \\ 0 & 0 & 0 & 0 & 0 & 0 & 0 & 0 & 0 & \sigma u_x \\ 0 & 0 & 0 & 0 & 0 & 0 & 0 & 0 & 0 & 0 \\ 0 & 0 & 0 & 0 & 0 & 0 & 0 & 0 & 0 & 0 \\ 0 & 0 & 0 & 0 & 0 & 0 & 0 & 0 & 0 & 0 \end{bmatrix}, \quad (A.16)$$

$$\mathbf{K}_{zy} = \begin{bmatrix} 0 & 0 & 0 & 0 & 0 & 0 & 0 & 0 & 0 & 0 \\ 0 & 0 & 0 & 0 & 0 & 0 & 0 & 0 & 0 & 0 \\ 0 & 0 & 0 & \mu & 0 & 0 & 0 & 0 & 0 & 0 \\ 0 & 0 & -\tfrac{2}{3}\mu & 0 & 0 & 0 & 0 & 0 & 0 & 0 \\ 0 & 0 & 0 & 0 & 0 & 0 & 0 & 0 & 0 & 0 \\ 0 & 0 & 0 & 0 & 0 & 0 & 0 & 0 & 0 & 0 \\ 0 & 0 & 0 & 0 & 0 & 0 & 0 & 0 & 0 & \sigma u_y \\ 0 & 0 & 0 & 0 & 0 & 0 & 0 & 0 & 0 & 0 \\ 0 & 0 & 0 & 0 & 0 & 0 & 0 & 0 & 0 & 0 \\ 0 & 0 & 0 & 0 & 0 & 0 & 0 & 0 & 0 & 0 \end{bmatrix}, \quad (A.17)$$

$$\mathbf{K}_{zz} = \begin{bmatrix} 0 & 0 & 0 & 0 & 0 & 0 & 0 & 0 & 0 & 0 \\ 0 & \mu & 0 & 0 & 0 & 0 & 0 & 0 & 0 & 0 \\ 0 & 0 & \mu & 0 & 0 & 0 & 0 & 0 & 0 & 0 \\ 0 & 0 & 0 & \tfrac{4}{3}\mu & 0 & 0 & 0 & 0 & 0 & 0 \\ 0 & 0 & 0 & 0 & \kappa_{hr} & 0 & 0 & 0 & 0 & 0 \\ 0 & 0 & 0 & 0 & 0 & \kappa_e & 0 & 0 & 0 & 0 \\ 0 & 0 & 0 & 0 & 0 & 0 & \sigma & -\sigma u_x & -\sigma u_y & 0 \\ 0 & 0 & 0 & 0 & 0 & 0 & 0 & 1 & 0 & 0 \\ 0 & 0 & 0 & 0 & 0 & 0 & 0 & 0 & 1 & 0 \\ 0 & 0 & 0 & 0 & 0 & 0 & 0 & 0 & 0 & 1 \end{bmatrix}, \quad (A.18)$$

$$\mathbf{S}_1 = \begin{bmatrix} 0 & 0 & 0 & 0 & 0 & 0 & 0 & 0 & 0 & 0 \\ 0 & 0 & 0 & 0 & 0 & 0 & 0 & 0 & 0 & 0 \\ 0 & 0 & 0 & 0 & 0 & 0 & 0 & 0 & 0 & 0 \\ 0 & 0 & 0 & 0 & 0 & 0 & 0 & 0 & 0 & 0 \\ 0 & 0 & 0 & 0 & -K_{eh} & K_{eh} & 0 & 0 & 0 & 0 \\ 0 & 0 & 0 & 0 & K_{eh} & -K_{eh} & 0 & 0 & 0 & 0 \\ 0 & 0 & 0 & 0 & 0 & 0 & 0 & 0 & 0 & 0 \\ 0 & 0 & 0 & 0 & 0 & 0 & 0 & 0 & 0 & 0 \\ 0 & 0 & 0 & 0 & 0 & 0 & 0 & 0 & 0 & 0 \\ 0 & 0 & 0 & 0 & 0 & 0 & 0 & 0 & 0 & 0 \end{bmatrix}, \quad (A.19)$$

$$\mathbf{S}_0 = \begin{bmatrix} 0 \\ J_{qy}B_z - J_{qz}B_y \\ J_{qz}B_x - J_{qx}B_z \\ J_{qx}B_y - J_{qy}B_x \\ \partial_t p_h + \mathbf{u}\cdot\nabla p_h \\ \partial_t p_e + \mathbf{u}\cdot\nabla p_e - 4\pi\varepsilon_r + \dfrac{5k_B}{2e}\mathbf{J}_q\cdot\nabla T_e + \mathbf{J}_q\cdot(\mathbf{E}+\mathbf{u}\times\mathbf{B}) \\ 0 \\ 0 \\ 0 \\ 0 \end{bmatrix}. \quad (A.20)$$